\documentclass[amsmath,amssymb,superscriptaddress,nobalancelastpage,prb,twocolumn,]{revtex4-1}
\AtBeginDocument{%
    \newwrite\bibnotes
    \def\bibnotesext{Notes.bib}
    \immediate\openout\bibnotes=\jobname\bibnotesext
    \immediate\write\bibnotes{@CONTROL{REVTEX41Control}}
    \immediate\write\bibnotes{@CONTROL{%
    apsrev41Control,author="08",editor="1",pages="1",title="0",year="1"}}
     \if@filesw
     \immediate\write\@auxout{\string\citation{apsrev41Control}}%
    \fi
}%

\usepackage{graphicx}
\usepackage{varioref}
\usepackage{xr-hyper}
\usepackage{xcolor}
\usepackage{nicefrac}
\usepackage{xfrac}
\usepackage{hyperref}
\hypersetup{colorlinks,linkcolor=blue,urlcolor=blue,citecolor=orange}
\usepackage{ulem}
\usepackage{siunitx}
\usepackage{graphicx}
\usepackage{dcolumn}
\usepackage{bm}
\usepackage{braket}
\usepackage{wasysym}
\usepackage{textcomp}
\usepackage{cancel}

\newcommand{\third}{{\ensuremath{\frac{1}{3}}}}

\begin{document}

\title{Intricacies of Frustrated Magnetism in the Kondo Metal YbAgGe}

\author{D.~G.~Mazzone}
\email{daniel.mazzone@psi.ch}
\affiliation{PSI Center for Neutron and Muon Sciences, 5232 Villigen PSI, Switzerland} 

\author{C.~B.~Larsen}
\affiliation{PSI Center for Neutron and Muon Sciences, 5232 Villigen PSI, Switzerland} 

\author{B. Ueland}
\affiliation{Department of Physics and Astronomy, Iowa State University, Ames, Iowa 50011, USA}
\affiliation{Ames National Laboratory, Iowa State University, Ames, Iowa 50011, USA}

\author{X.~Boraley}
\affiliation{PSI Center for Neutron and Muon Sciences, 5232 Villigen PSI, Switzerland} 

\author{D. M. Pajerowski}
\affiliation{Neutron Scattering Division, Oak Ridge National Laboratory, Oak Ridge, Tennessee 37831, USA}

\author{Y. Skourski}
\affiliation{Dresden High Magnetic Field Laboratory (HLD-EMFL),
Helmholtz-Zentrum Dresden-Rossendorf, 01328 Dresden, Germany}

\author{J. Taylor}
\affiliation{ISIS Facility, STFC Rutherford Appleton Laboratory,
Harwell Science and Innovation Campus, Oxfordshire OX11 0QX, United Kingdom}
\affiliation{Neutron Scattering Division, Oak Ridge National Laboratory, Oak Ridge, Tennessee 37831, USA}

\author{B.~F\aa k}
\affiliation{Institut Laue-Langevin, 71 Avenue des Martyrs, CS20156, 38042 Grenoble C\'edex 9, France}


\author{S. L. Bud’ko}
\affiliation{Department of Physics and Astronomy, Iowa State University, Ames, Iowa 50011, USA}
\affiliation{Ames National Laboratory, Iowa State University, Ames, Iowa 50011, USA}

\author{R. McQueeney}
\affiliation{Department of Physics and Astronomy, Iowa State University, Ames, Iowa 50011, USA}
\affiliation{Ames National Laboratory, Iowa State University, Ames, Iowa 50011, USA}

\author{P. C. Canfield}
\affiliation{Department of Physics and Astronomy, Iowa State University, Ames, Iowa 50011, USA}
\affiliation{Ames National Laboratory, Iowa State University, Ames, Iowa 50011, USA}

\author{O.~Zaharko}
\email{oksana.zaharko@psi.ch}
\affiliation{PSI Center for Neutron and Muon Sciences, 5232 Villigen PSI, Switzerland}

\date{\today}
             
\begin{abstract}

\end{abstract}

   \maketitle

\textbf{The combination of localized magnetic moments, their frustration and interaction with itinerant electrons is a key challenge of condensed matter physics. Frustrated magnetic interactions promote degenerate ground states with enhanced fluctuations, a topic that is predominantly studied in magnetic insulators. The coupling between itinerant and localized electrons in metals add complexity to the problem, and is presently formulated only for extreme cases in which the itinerant electrons mediate exchange between localized spins (RKKY interaction) or suppress the formation of magnetic moments (Kondo screening). Here, we report an in-depth experimental study of the distorted Kagome metal YbAgGe, unravelling the open questions of how frustration, localized magnetism and itinerant electrons are intertwined in frustrated Kondo metals. We find that coupled itinerant and localized electrons give rise to dynamic magnetic correlations below $T^* \approx$ 20 K. At lower temperature, frustrated magnetic interactions establish anisotropic magnetic short-range correlations that culminate into antiferromagnetic long-range order below $T_N$ = 0.68 K with a significantly reduced modulated magnetic moment. We show that local moment Hamiltonians can yield limited understanding of the microscopic behaviour in frustrated metals, and prompt the extension of more sophisticated model Hamiltonians incorporating itinerant effects.}

\section*{Introduction}

Magnetic frustration in materials arises from constraints of the underlying lattice geometry or conflicting exchange interactions, which can yield degenerate magnetic configurations with unusual ground states such as skyrmion lattices\cite{Fert2017}, spin glasses \cite{Mydosh1993}, 
spin liquids\cite{Balents2010} and other phases that may be attractive for future technological applications. A microscopic understanding of these frustrated magnetic states is often achieved through local-moment magnetic Hamiltonians with nearest-neighbour spin interactions \cite{Starykh2015,broholm2020}. These microscopic models have proven fruitful for insulating materials in which itinerant electrons can be neglected, further neighbour interactions decay promptly with distance, and higher-order spin interactions are strongly suppressed \cite{Hayami2021}.
The approach is less effective for magnetically frustrated metals. For this material class it is often  necessary to account for long-range Fermi surface effects, in which conduction electrons interact with the local moments of the magnetic ions.

Microscopic Hamiltonians appropriate for magnetic metals with embedded localized moments are derived from the Kondo lattice Hamiltonian (KLH) $H_{KLH}$ = $H_{kin}$ + $H_{ex}$\cite{Hayami2021,Hassanieh2010}. Here, $H_{kin}$ describes the kinetic motion of the itinerant electrons, and $H_{ex}$ represents the coupling of the localized spins with the conduction electrons. The KLH is particularly important for rare-earth compounds \cite{Stewart2001}, for which the relevant RKKY interaction \cite{Ruderman1954,Kasuya1956,Yosida1957} can be derived as lowest-order contribution \cite{Hayami2021}. In this model the conduction electrons serve as exchange pathways between local magnetic moments, directly encoding the Fermi surface into local exchange parameters $J$ of effective Heisenberg-like Hamiltonians. Within this approach, it is possible to treat magnetism in metals similarly to the insulating case, but with anisotropic exchanges interactions. This allows for the exploration of novel frustrated quantum phases in metals \cite{Hayami2021,Lacroix2010,Stockert2020,Kurumaji2019,Gao2016}, though their interpretation often remains challenging.   

\begin{figure*}[tbh]
\centering
\includegraphics[width={\textwidth}]{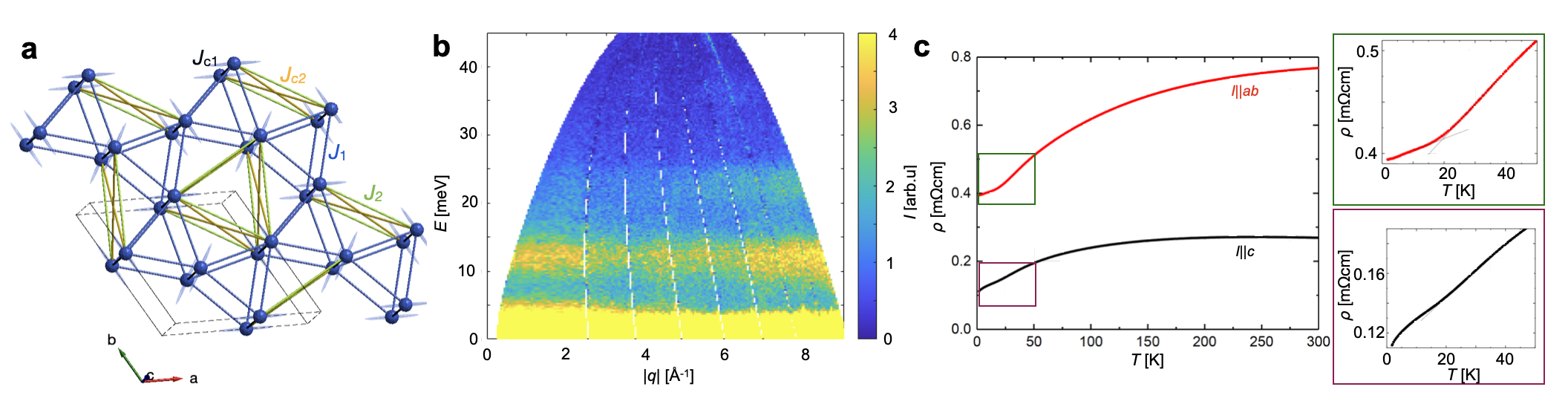}\\
\caption{\label{fig:CSMES}%
\textbf{The three main ingredients in frustrated Kondo metals.} \textbf{a} \textit{Frustration:} the Yb$^{3+}$ ions in YbAgGe generate a distorted Kagome structure with nearest and next-nearest neighbour interactions $J_1$ (blue) and $J_2$ (green) in the hexagonal plane and nearest neighbour $J_{c1}$ (black) and $J_{c2}$ (gold) interplane interactions. 
Single ion anisotropy ellipsoids are shown in blue. \textbf{b} \textit{Local magnetism:} inelastic neutron scattering data for energy transfers $E<$ 45 meV and wavector transfers $|q|<$ 9 \AA$^{-1}$ measured at $T$ = 3 K on the MARI spectrometer. The transition at $E$ = 12 meV originates from a CEF excitation mixed with a phonon, the transition at $E$ = 23 meV is attributed to phononic contributions. \textbf{c} \textit{Itinerant electrons:} electrical resistivity measurements in which the electric current $I$ runs parallel or perpendicular to the crystallographic $c$-axis. Below $T^*$ = 20 K additional scattering processes yield deviations from the temperature dependence at higher temperature. Note that the material is known to possess a high residual resistivity value that does not arise from crystalline disorder, but stems from scattering processes which can be suppressed efficiently with magnetic field \cite{Budko2004}. }
\end{figure*}

Metals containing rare-earth ions such as Ce, Eu, Sm and Yb experience an additional interaction. The conduction electrons can (partially) screen localized moments, engaging them into collective singlet states \cite{Kondo1964}. This Kondo screening process can be understood through an extension of $H_{ex}$, encrypting an antiferromagnetic coupling between the localized $f$-electrons and the conduction bands \cite{hewson1997kondo}. The admixture between these electrons can become coherent at low temperature, leading to hybridized electron bands with heavy effective masses close to the Fermi surface. A microscopic band description of these heavy-fermion systems has proven to be notoriously challenging, as most $ab$-$initio$ calculations have difficulties to distinguish the small energy differences of the various emergent states. Magnetic frustration complicates the matter, because theoretical predictions need to ascertain whether various spin configurations are degenerate or separated in energy. To this end, we address the question of how the three ingredients - RKKY interaction, Kondo screening and frustration - intertwine to establish the physics in YbAgGe.  

\section*{Results}

\subsection*{Local Magnetism and Frustration}

YbAgGe is a member of the RAgGe family crystallizing in the hexagonal ZrNiAl-type crystal structure (space group $P\bar{6}2m$)\cite{Gibson1996}, in which Yb$^{3+}$ ions form distorted Kagome layers with periodicity $a$ = $b$ = 7.05 \AA~perpendicular to $c$ = 4.14 \AA~(see Fig. \ref{fig:CSMES}a and Supplementary Information (SI) Note 1 for details). Within the localized model the overarching magnetic energy scale is determined by the localized Yb$^{3+}$ Kramers ions, with total angular momentum $J$ = 7/2. The 2$J$+1 = 8 degenerate eigenstates of the Yb$^{3+}$ ion are split by the non-spherical crystal-electric field (CEF). The lowest CEF transition has at an energy transfer of $E$ = 12.0(4) meV, manifesting as a non-dispersing excitation in our polycrystalline inelastic neutron scattering (INS) experiment on the MARI spectrometer at ISIS (Fig. \ref{fig:CSMES}b). The transition is in agreement with earlier reports \cite{Matsumura2004,Bonville2007}. Using PyCrystalField\cite{scheie2021} our calculation establishes that the ground state retains the axial anisotropy, aligned with 2-fold axes passing through the Yb$^{3+}$ ions (see Fig. \ref{fig:CSMES}a) and predicts a magnetic moment of $\mu_{CEF}$ = 3.98$\mu_B$ (see SI Note 2 for details).

\begin{figure*}[tbh]
\centering
\includegraphics[width=\textwidth]{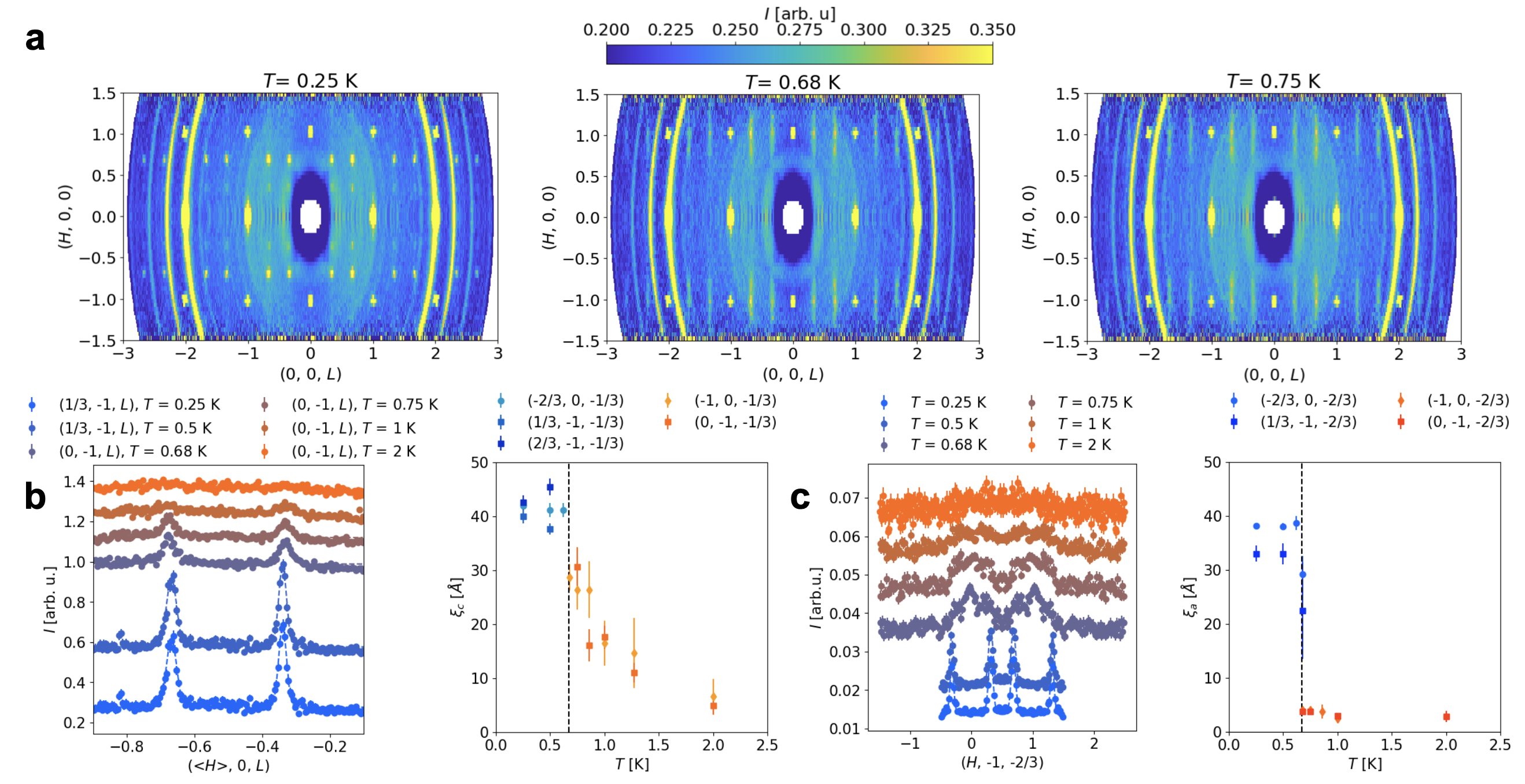}\\
\caption{\label{fig:diffraction_1}%
\textbf{Static magnetism in YbAgGe.} \textbf{a} Two dimensional reciprocal space map in the ($H$, 0, $L$) plane measured at $T$ = 0.25, 0.68 and 0.75 K. Below $T \approx$ 2 K antiferromagnetic short-range correlations emerge at a wavevector $q$ = (0, 0, {\third}) with a correlation length along the $c$-axis which exceeds the in-plane correlation length. Below a first-order transition at $T_N$ = 0.68 K long-range order is developed at  $q $ = ({\third}, 0, {\third}). The data were measured at the Cold Neutron Chopper Spectrometer (CNCS) using an incident energy of $E_i$ = 12 meV. The figures were generated using an energy integration $\Delta E$ = [-1/2, 1/2] meV and a reciprocal space integration $\Delta Q$ = [-0.1, 0.1] rlu perpendicular to the reciprocal plane. \textbf{b} One dimensional line cuts and quantitative analysis of the short-range correlations in YbAgGe along the crystallographic $c$- and \textbf{c} $a$-direction. The four dimensional data was integrated over $\Delta E$ =$\pm0.25$ meV around the elastic line and $\Delta Q$ =$\pm0.05$ rlu perpendicular to the shown cutting direction. The cuts are offset for clarity. The resulting one dimensional line-cuts were fitted by Gaussian line shapes, from which the correlation lengths $\xi_c$ = $c$/$\pi$FWHM and $\xi_a$ = $a$/$\pi$FWHM were determined as function of temperature. FWHM denotes the full-width at half-maximum of the Gaussian peaks.}
\end{figure*}

\begin{figure*}[tbh]
\centering
\includegraphics[width=0.9\textwidth]{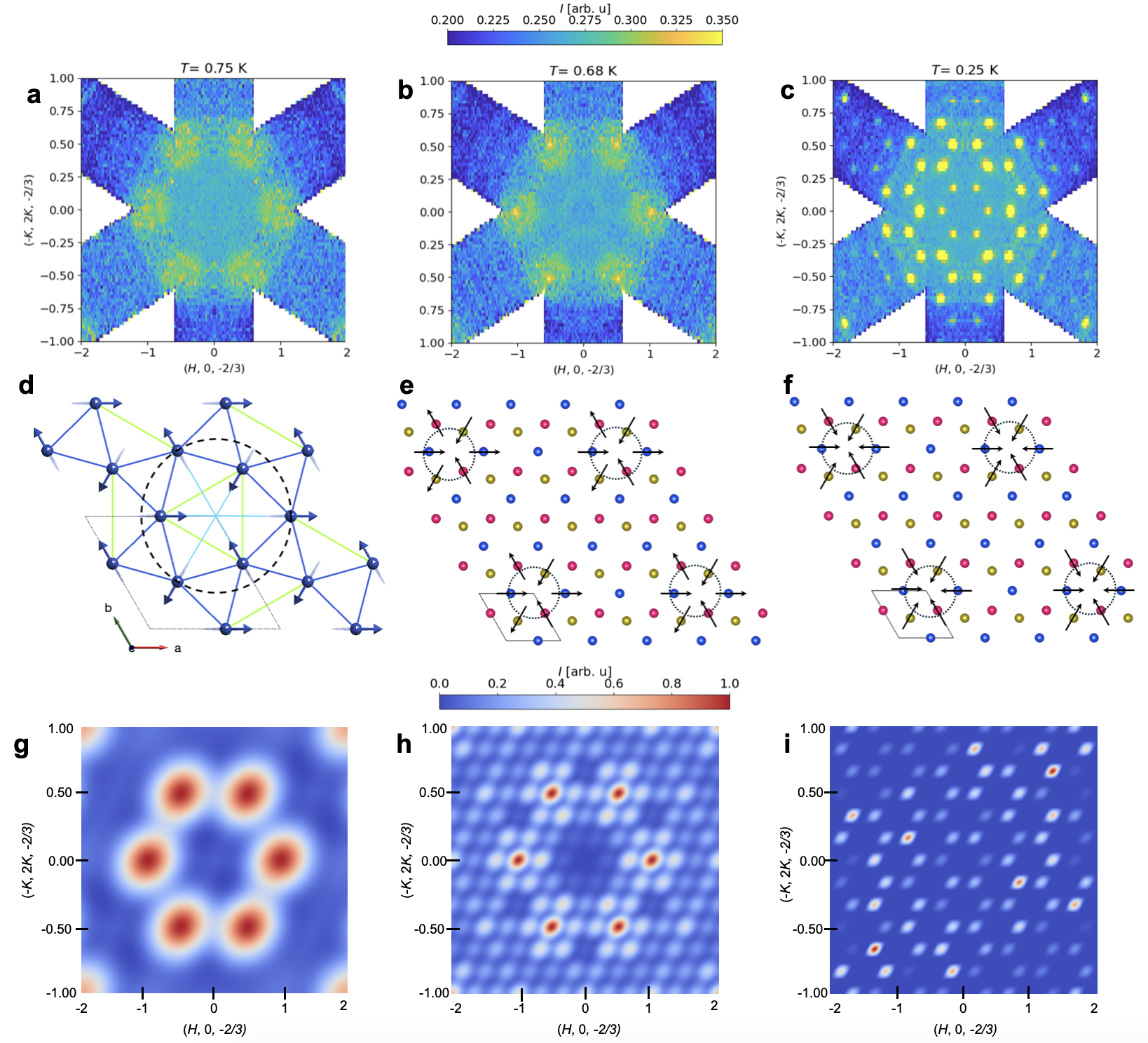}\\
\caption{\label{fig:diffraction_2}%
\textbf{Static magnetism in YbAgGe.} Two dimensional reciprocal space maps perpendicular to (0, 0, -2/3) measured at $T$ = 0.75 \textbf{a}, 0.68 \textbf{b} and 0.25 K \textbf{c}. The data were measured at CNCS using an incident energy of $E_i$ = 12 meV. The figures were generated using an energy integration $\Delta E$ = [-1/2, 1/2] meV and a reciprocal space integration $\Delta Q$ = [-0.69, -0.63] rlu perpendicular to the reciprocal plane.  \textbf{d} The magnetic moment arrangement within the hypothetical $q$ = (0, 0) structure, a dashed black circle spans over a Yb-hexagon. \textbf{e} The 3$a^*$ - 3$b^*$ structure with the parallel moment alignment connected by $J_{3a}$ (labeled in Supplementary Fig. S3b). \textbf{f} The 3$a^*$ - 3$b^*$ structure with the antiparallel moment alignments.
Calculated ($H$, $K$) cuts simulated via Fourier transform of \textbf{g} a 3x3x$N$ box containing the hypothetic $q$ = (0, 0) structure, \textbf{h} a 10x10x$N$ box with the parallel moment alignment connected by $J_{3a}$, \textbf{i} a 24x24x$N$ box containing three 120 deg rotated domains of the zero-field $q_1$=({\third},~0,~{\third}) amplitude modulated structure \cite{Larsen2021} with parallel and antiparallel moment alignments connected by $J_{3a}$.}
\end{figure*}

Within the localized picture the Heisenberg exchange Hamiltonian approximation enables to predict effective magnetic models for the $R$AgGe crystal geometry. They range from Kagome spin layer models, when the frustration between intraplane interactions $J_{i}$ prevails, to spin chain models, when the frustration is dominant within the interplane interactions $J_{ci}$. For instance, in the $R$ = Tm member the energy hierarchy can be resolved as follows:\cite{Larsen2023} 
i) the CEF-anisotropy dominates the magnetic exchange interactions; 
ii) the strongest magnetic exchange coupling $J_{c}$ along the $c$-axis is ferromagnetic (FM) and not frustrated; 
iii) the interplay between weaker, intraplane $J_{1}$, $J_{2}$ couplings manifests ice rules for the $J_{2}$ triangles, but the spin-ice frustration is released by the significant equidistant but distinct $J_{3a}$ and $J_{3b}$ terms (see also SI Note 3). In YbAgGe the CEF-anisotropy is the same as in TmAgGe, but its contribution is  less strong. This allows for interplane interactions $J_{ci}$ to gain in importance yielding frustration effects alongside the interplane couplings. 

Macroscopic studies and earlier neutron diffraction results  
\cite{Larsen2021} have reported that antiferromagnetic (AFM) long-range order (LRO) in YbAgGe is established below $T_N$ = 0.68 K with a propagation vector of $q$ = ({\third},~0,~{\third}). The order is unconventional, featuring varying moment sizes on different atoms with an average amplitude of 1.62(3)$\mu_B$/Yb. A magnetic field applied in the Kagome plane triggers a series of metamagnetic transitions to other LRO states with different periodicity in the Kagome plane, but with persistent $q_z \approx$ {\third} component. Above  $\mu_0H\approx$ 7 T ($H\parallel$ [1, -1, 0]) these modulations vanish, so that eventually only an induced FM component remains (see SI Note 1 for details). We attempted to conceive the dominant exchange interactions in YbAgGe by minimizing the Heisenberg-type Hamiltoninan and analyzing the solutions with the propagation vectors determined in earlier experiments\cite{Larsen2021} (see SI Note 3). We find that the $q$ = ({\third},~0,~{\third}) propagation vector requires AFM nearest-neighbour interplane interactions $J_{c1}$ and $J_{c2}$, and finite further neighbour intraplane terms $J_3$. 

Previous thermodynamic measurements have shown that YbAgGe
exhibits increased magnetic correlations already above $T_N$, for $T <$ $T_{SR}$ $\approx$ 2 K \cite{Budko2004}. Here we reveal their microscopic origin via single crystal elastic scattering maps, which we measured on the Cold Neutron Chopper Spectrometer (CNCS) at ORNL. Figures \ref{fig:diffraction_1}a and \ref{fig:diffraction_2}a show elastic scattering maps in the ($H$, 0,  $L$) and ($H$, $K$, -2/3) planes for $T$ = 0.25, 0.68 and 0.75 K. Upon cooling below $T$ $\lesssim$ 2 K diffuse scattering develops  around the $q$ = (0, 0, {\third}) positions (see Fig. \ref{fig:diffraction_1}). The correlation lengths $\xi_c$ and $\xi_a$ show that at $T$ = 2 K the material is correlated over $\sim$5 \AA~ along the $c$-direction, and $\sim$4 \AA~ along the $a$ direction. Upon cooling $\xi_c$($T$) is continuously increasing  and reaches the resolution limited correlation of structural Bragg peaks ($\sim$10 unit cells) below $T$ = 0.68 K where magnetic long-range order with $q$ = ({\third},~0,~{\third}) is established. This behavior is different for $\xi_a$($T$), where the correlation length remains unchanged above $T$ $>$ 0.68 K. A discontinuous jump of $\xi_a$($T$) is observed at $T$ = 0.68 K below which the correlation length reaches 5 unit cells matching the correlation length of structural Bragg peaks.

The diffuse scattering patterns correspond to spin-spin correlations that were modeled using fast Fourier transform and sampling theory to reduce the high-frequency noise in the calculations as implemented by the Scatty program \cite{Paddison2018}. Above $T_N$ = 0.68 K YbAgGe is short-range ordered within the hexagonal plane but reveals extended correlations along the $c$-axis. Thus, we neglected the $J_{cij}$ terms (see Fig. \ref{fig:CSMES}a) because the $q_z$= {\third} correlations are already established at these temperatures. We first modeled the diffuse scattering at $T$ = 0.75 K, for which we used a hypothetical two-dimensional long-range $q$ = (0, 0) structure with 120 deg canted magnetic moments such as presented in (Fig. \ref{fig:diffraction_2}d). 
The order represents a compromise between the local CEF anisotropy and AFM $J_1$ and $J_2$ couplings; the moments connected by the $J_{3a}$- and $J_{3b}$- bonds are parallel within such an arrangement. The corresponding reciprocal $hk$-plane consist of sharp magnetic Bragg peaks at integer positions with $h+k=2n+1$. Magnetic correlations confined to a few unit cells result in diffuse scattering around these $hk$ positions, which is reproduced by a small number of unit cells along the $a$- and $b$-axes. Notably, the Fourier transform of a box with 3x3x$N$ unit cells qualitatively reproduces the diffuse pattern measured at $T$ = 0.75 K (Fig. \ref{fig:diffraction_2}g). The good agreement is interpreted as evidence of prevalent $q$ = (0, 0) correlations extending over three unit cells. They result from the local anisotropy and AF $J_1$ and $J_2$ couplings. We attempted to reproduce the pattern at $T$ =  0.68 K by i) considering parallel moment alignments on the $J_{3b}$ bonds, ii) populating only every third hexagon with a magnetic moment and suppressing the moment on other sides as shown in Fig.~\ref{fig:diffraction_2}e, and iii) limiting the size of coherent spin arrangements to 10 unit cells along the  $a$- and $b$-axis. Figure~\ref{fig:diffraction_2}h shows the Fourier transform of a box containing 10x10x$N$ unit cells, confirming the success of this approach. The model captures the main features of the measured $T$ = 0.68 K pattern, $i.e.$ the intensity at the main $hk$ positions and the smeared satellites at $q$ = ({\third}, 0) and its equivalents. At $T$ = 0.25 K the ordering vector $q$ = ({\third},~0,~{\third}) and its arms supersede any remains of the $q$ = (0, 0) structure (see Fig.~\ref{fig:diffraction_2}c). We can reproduce the diffraction pattern (see Fig.~\ref{fig:diffraction_2}i) by Fourier transforming a 24x24x$N$ box containing three domains with the ({\third},~0),({0,~\third}) and ({-\third},~{\third}) propagation vectors and the amplitude modulated arrangement proposed in Ref.\cite{Larsen2021}. In this model  $J_{3a}$ and $J_{3b}$ connect alternating parallel and antiparallel moment alignments, as shown in Fig. \ref{fig:diffraction_2}f.
Our results suggest that the long-range magnetic order at low temperature is established  as a compromise between the CEF anisotropy, AFM $J_1$ and $J_2$ couplings and competing AFM $J_{3a}$ and $J_{3b}$ couplings.

\subsection*{Itinerant electrons}

The metallic nature of the $R$AgGe family adds complexity to understanding the magnetism. The itinerant electrons affect the degeneracy of frustrated states and can lead to significant further neighbour exchange couplings and Kondo screening. There exist a number of experimental evidences in YbAgGe, which do not comply with a localized moment picture. For instance, below $T^* \approx$ 20 K, the temperature at which the full $R$ln2 entropy is recovered \cite{Budko2004, Morosan2004}, our electric resistivity measurements with currents $I$ parallel and perpendicular to the $c$-axis show deviations to the behavior at high temperature (see Fig. \ref{fig:CSMES}c). These data, together with previous magnetization and specific heat results reporting a stark deviation from the Curie-Weiss behavior below $T^*$ and a Sommerfeld coefficient between 150 mJ/molK$^2$ $\le$ $\gamma$  $\le$ 1200 mJ/molK$^2$  \cite{Budko2004, Morosan2004} indicate a link between conduction electrons and local magnetic degrees of freedom, suggesting hybridization of Yb 4$f$ electrons with the conduction bands. Strong support for Kondo physics is also obtained from earlier neutron scattering results, reporting a substantial broadening of the quasielastic scattering and the first excited crystal-field level as a function of temperature \cite{Matsumura2004,Fak2005}. Besides, our magnetization measured at high magnetic fields is incompatible with a localized picture (see SI Note 4). Another argument for itineracy is the reduced and amplitude modulated ordered magnetic moment value as determined from neutron diffraction. Even inside the AFM LRO state at zero field,  
the ordered moment is only $\mu_{ord}\approx$  1.6$\mu_B$ on average\cite{Larsen2021} implying a loss of $\sim$60\% when compared to the expected localized moment $\mu_{CEF}$ = 3.98$\mu_B$.

\begin{figure*}[tbh]
\centering
\includegraphics[width=0.9\textwidth]{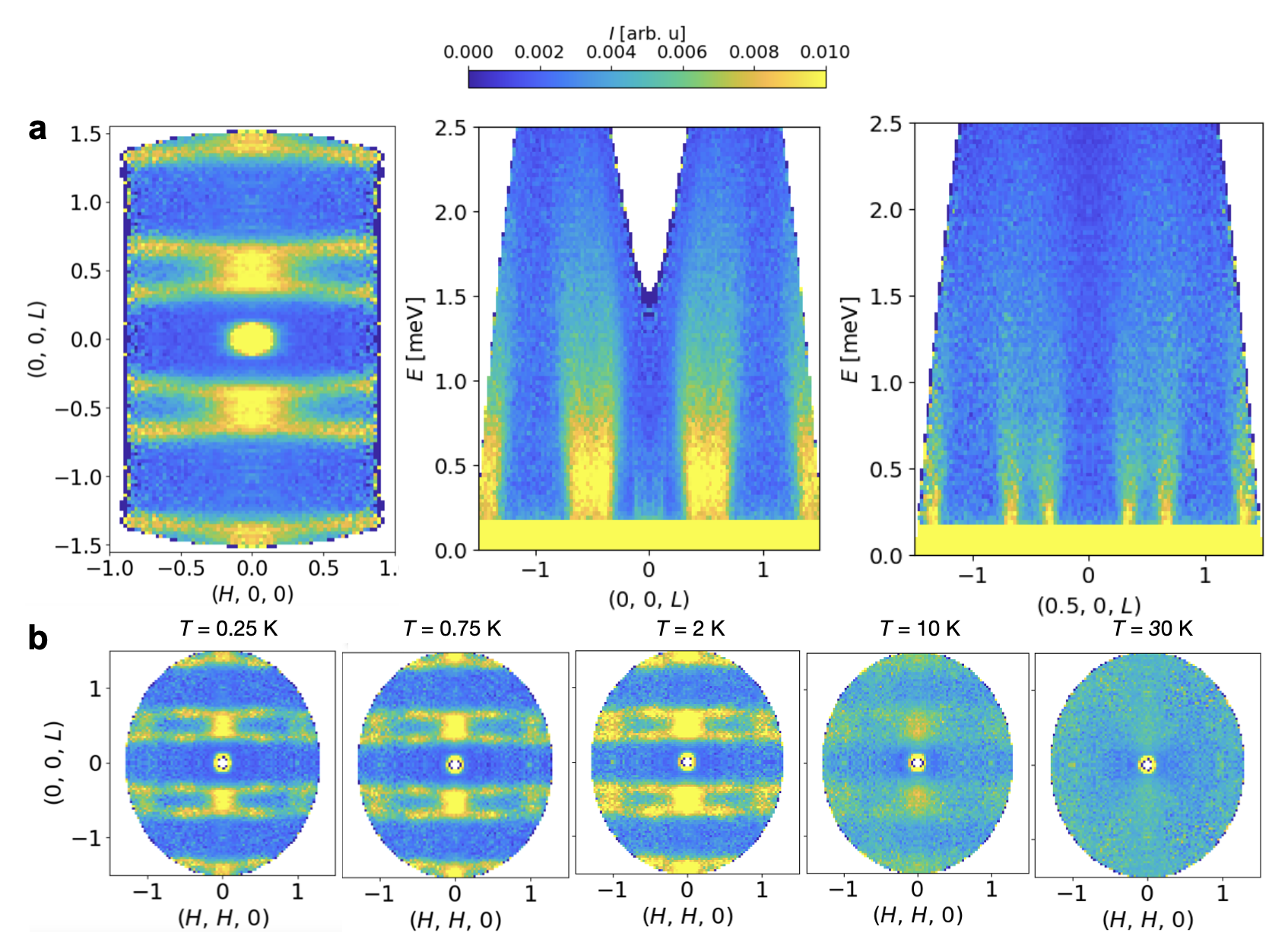}\\
\caption{\label{fig:inelastic}%
\textbf{Dynamic properties of YbAgGe.} \textbf{a} Magnetic excitation spectrum at $T$ = 0.25 K and $\mu_0H$ = 0 T in the ($H$, 0, $L$) plane at an energy transfer $E$ = 0.3 meV, along (0, 0, $L$) and (1/2, 0, $L$). The data were measured on CNCS using an incident energy $E_i$ = 3.32 meV and plotted with an energy integration $\Delta E$ = [0.2, 0.4] meV and reciprocal space integration $\Delta Q$ = [-0.1, 0.1] rlu in the first subpanel, and $\Delta Q$ = $\pm0.1$ rlu perpendicular to the (0, 0, $L$) and (1/2, 0, $L$) directions. \textbf{b} Reciprocal space maps of the ($H$, $H$, $L$) plane at an energy transfer $E$ = 0.3 meV at $T$ = 0.25, 0.75, 2, 10 and 30 K. The data were measured with $E_i$ = 3.32 meV using an energy integration range $\Delta E$ = [0.2, 0.4] meV and $\Delta Q$ = [-0.1, 0.1] rlu perpendicular to the ($H$, $H$, $L$) plane.
}
\end{figure*}

Our most striking counter to the localized picture is provided by our inelastic neutron scattering results. An analysis of the dipole-allowed transitions between the CEF ground state wavefunctions predicts the absence of observable magnetic excitations (see SI Note 2 for details). This is in strong contrast to our experimental findings shown in Fig. \ref{fig:inelastic}a measured at $T$ = 0.25 K and $\mu_0H$ = 0 T on CNCS. We observe enhanced spectral weight along ($H$, 0, 1/3), ($H$, 0, 2/3) and around (0, 0, 1/2). The intensity is modulated mainly along the $c$-axis, but the spectrum also features an anisotropy in the Kagome plane. The excitations are column-like with decreasing intensity as function of energy transfer. No clear excitation gap is observed, instead we find that the excitation spectrum resembles an energy dependence reported previously for heavy-fermion metals \cite{Amato1988,Kadowaki2004,Knafo2004}. In accordance to these results and an earlier report on YbAgGe \cite{Fak2005} we fitted the dynamic susceptibility with a quasielastic Lorentzian, and find that the function adequately describes our experimental results (see SI Note 5 for details).

\begin{figure}[tbh]
\centering
\includegraphics[width=\linewidth]{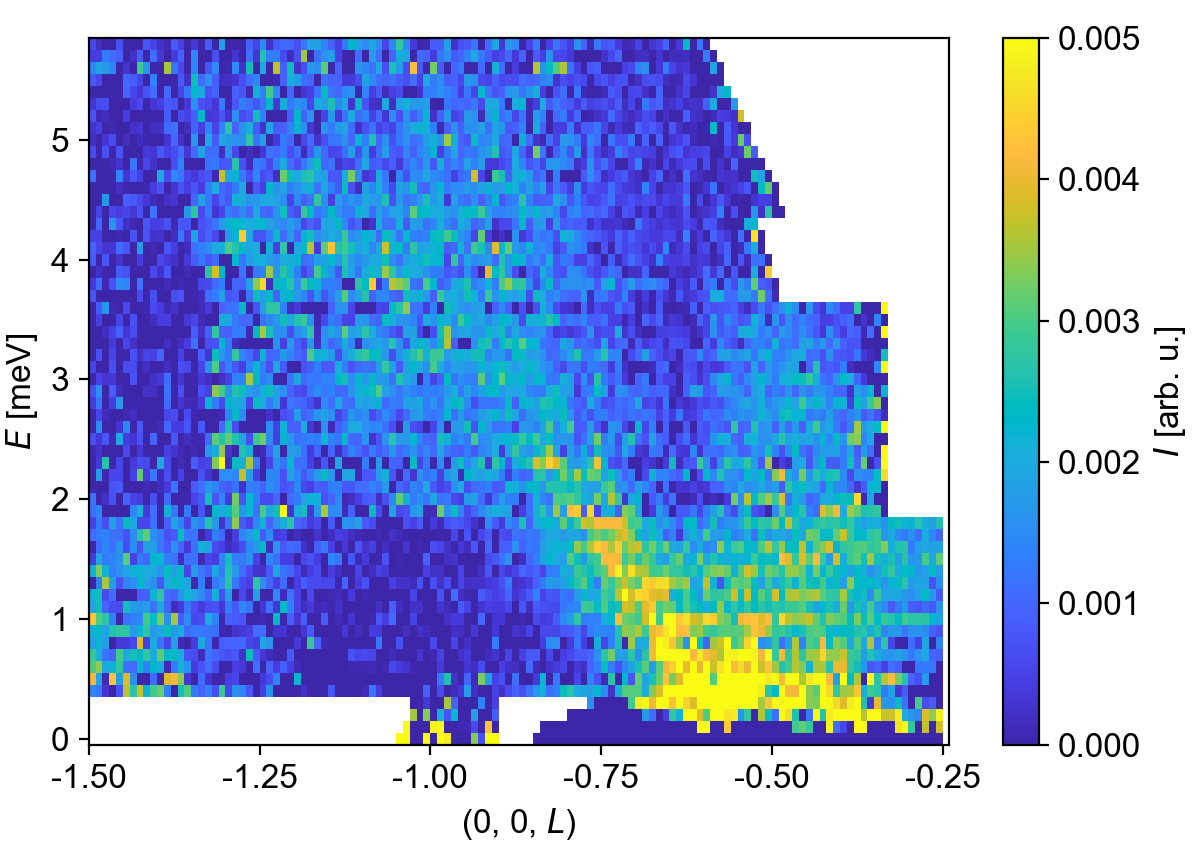}\\
\caption{\label{fig:INSfield}%
\textbf{Dynamic properties in magnetic field.} Magnetic excitation spectrum at $T$ = 0.05 K and $\mu_0H$ = 11 T along (0, 0, $L$). The background subtracted data were measured at CAMEA with a reciprocal space integration $\Delta Q$ = [-0.25, 0.25] rlu along the ($H$, $H$, 0) direction.}
\end{figure}

Remarkably, the magnetic excitation spectrum is not affected by moderate temperature changes. In Fig. \ref{fig:inelastic}b we display reciprocal space maps of the ($H$, $H$, $L$) plane for $E$ = 0.3 meV measured at $T$ = 0.25, 0.75, 2, 10 and 30 K. The results imply that magnetic fluctuations are unaffected by magnetic long-range order at $T_N$ = 0.68 K. Notably, they are gradually weakened and suppressed only above $T^* \approx$ 20 K (see SI Note 5 for details), matching the unusual temperature dependence of the electronic resistivity (see Fig. \ref{fig:CSMES}c). The INS spectra also reveal a nontrivial behaviour under magnetic field. Our results measured at CAMEA with $T$ = 50 mK and $H\parallel$ [1, -1, 0] show that the magnetic fluctuations are only modified above $\mu_0H \geq$ 7 T where magnetic long-range order with $q_z\approx$ {\third} is suppressed (see SI Note 6 and Ref. \cite{Larsen2021}). In fact, at $\mu_0H$ = 11 T we find that the excitations develop a dispersion along (0, 0, $L$) that is quasielastic at (0, 0, 1/3) and (0, 0, 2/3) and reaches a maximum at (0, 0, 1) with an energy transfer $E$ $\sim$3.5 meV (see Fig. \ref{fig:INSfield}). The full-width at half-maximum of the magnetic excitation on top of the dispersion is  about three times larger than the instrumental resolution $i.e.$ $\Delta E$ = 1.03(8) meV, providing evidence for a finite excitation lifetime.

We tested a variety of minimal local-moment Hamiltonians that retain $q_z\approx$ {\third} to reproduce the magnetic excitation spectrum along (0, 0, $L$). Notably, we considered one-dimensional Heisenberg chains with an axial CEF anisotropy, anisotropic $xxz$-exchange interactions allowed by the crystal structure, and tested the effect of a magnetic field along $H\parallel$ [1, -1, 0] (see SI Note 7 for details). As expected, we find that even a small anisotropy induces a finite gap in the excitation spectrum, contrasting the experimental observations at zero field and under magnetic field. This further supports that the microscopic nature of the magnetic properties in YbAgGe go beyond the localized-moment picture, and indicates that magnetism is mediated by itinerant electrons.

\section*{Discussion}

YbAgGe is an intricate system showing elements of magnetic frustration, local moment and itinerant electrons physics. Our diffuse neutron scattering results shown in Fig. \ref{fig:diffraction_2}a highlight that magnetic frustration stemming from the distorted Kagome plane and from orthogonal chain interactions plays an important role in the material. Localised exchange Hamiltonians are the standard approach for insulators and are also appropriate for metals possessing only a small density of states at the Fermi surface \cite{Boraley2025}. However, this approach is inappropriate for metals like YbAgGe, because the strong interactions between local magnetic moments and itinerant electrons trigger preeminent hybrid excitations fainted by substantial damping effects \cite{Halloran2025,Chen2020,ROSSATMIGNOD1988376,Riberolles2024,Riberolles2023,Riberolles2022}. One such example with qualitative similarities to YbAgGe is the frustrated Kagome lattice TbMn$_6$Sn$_6$ \cite{Riberolles2024,Riberolles2023,Riberolles2022}. The material features sharp acoustic magnons and overdamped flat-band excitations that induce in-plane structure factor patterns broadly resembling the diffuse scattering we observe in YbAgGe above $T_N$ (see Fig. \ref{fig:diffraction_2}). The excitations in TbMn$_6$Sn$_6$ originate from flat-bands and chiral fluctuations localised on hexagonal plaquettes. However, since the ordered magnetic structure of YbAgGe is amplitude modulated and the moment directions are pinned by the CEF anisotropy, these short-range
correlations are possibly governed by different fluctuations.

Most magnets with prominent itineracy require  magnetic Hamiltonians which account for the electronic-band structure \cite{Goremychkin2018,Simeth2023,Ghioldi2024}. However, the small energy differences between different electronic configurations in Kondo metals still cause severe difficulties in theoretically predicting the electronic properties with sufficient accuracy, which complicates a mapping onto reliable effective Hamiltonians. Progress has been made in model cases of Kondo metals crystallizing  in simple cubic lattice structures such as CeX$_3$ (X = Pd, In)\cite{Goremychkin2018,Simeth2023,Ghioldi2024}. Here microscopic theoretical predictions have become possible, but their extension to complex structures such as non-centrosymmetric YbAgGe is not yet available. This step is required to quantitatively assess the effect of magnetic frustration in this material class. In fact, an intermediate route succeeding with this endeavor may include theoretical investigations of itinerant frustrations hosting Kondo inactive 4$f$-ions. Examples include other materials in the $R$AgGe family, also revealing unusual frustrated quantum phases without the additional complexities imposed by the Kondo exchange interaction \cite{Larsen2023, Kan2020, Bhandari2024,Schwertfeger2025}. Such studies  are important to deepen our understanding of how frustration is established in metallic hosts.

On a qualitative level our various experimental observations can be grouped into four different temperature regions. Above $T$ $>$ $T^*\approx$ 20 K YbAgGe behaves like a paramagnetic intermetallic material. As the temperature is decreased below $T^*$ a significant coupling between the itinerant electrons and the local magnetic moments is evidenced by the coincident change in the resistivity versus temperature behaviour and the emergence of spin-spin correlations in the INS data (see Figs. \ref{fig:CSMES}c and \ref{fig:inelastic}). The similarity to other Kondo metals \cite{Goremychkin2018,Amato1988,Kadowaki2004,Knafo2004} suggests that the coupling in YbAgGe arises via Kondo screening leading to hybridized Yb-bands with coherent inter-band excitations. If so, the field dependence of the magnetic excitations further suggests that the hybridized Yb-bands are located in the eminent vicinity of the Fermi surface, because they are affected at moderate fields already. Confirmation of this hypothesis requires detailed knowledge of the Fermi surface topology, which can be gained using angle-resolved photoemission spectroscopy, scanning tunneling microscope spectroscopy or De Haas-Van Alphen techniques. Unfortunately, technical challenges have restricted success of these methods in YbAgGe so far. 

Below $T_{SR}$ $\approx$ 2 K macroscopic heat capacity and electric resistivity measurement have shown that  the magnetic properties change \cite{Budko2004}. For 0.68 K $<$ $T$ $<$ $T_{SR}$ we find evidence for magnetic short-range correlations centered around $q$ = (0, 0, 1/3), and long-range along the $c$-direction (see Fig. \ref{fig:diffraction_1} and \ref{fig:diffraction_2}). The liquid-like state reveals similarities to a smectic phase with one long-range correlated axis only. While future theoretical efforts are required to elucidate the microscopic nature of this putative spin liquid-like phase, we find that it reveals growing correlations along the $c$-direction when the temperature is lowered while correlations in the $ab$-plane remain  short-ranged (see Fig. \ref{fig:diffraction_1}b and c). Using the localized Heisenberg picture our analysis suggests AFM $J_1$ and $J_2$ couplings in the Kagome plane, and equal sized $J_{c1}$ and $J_{c2}$ interplane couplings to realize $q$ = (0, 0, 1/3). 

Below $T_N$ = 0.68 K magnetic long-range order is established along $q$ = (1/3, 0, 1/3), in which in-plane ordering requires AFM $J_{c1}$ and $J_{c2}$ couplings and finite further neighbour terms in a localized picture (see SI Note 3). The diffraction results \cite{Larsen2021} report a significantly-reduced magnetic moment of $\mu_{ord}\approx$  1.6$\mu_B$ with respect to the expected localized moment of $\mu_{CEF}$ = 3.98$\mu_B$ obtained from the CEF ground state wavefunction. These observations ascertain that a localized picture provides only a limited view, so that the itineracy plays an important role. This may be reflected also in the stability of $q_z\approx$ {\third}, which can be interpreted as robust Fermi surface nesting along the $c$-axis, resisting changes induced by an applied magnetic field. Along these lines the field modifications in $q_x$ and $q_y$ indicate almost degenerate nesting conditions in the hexagonal plane that can be tuned with magnetic field. The hypothesis is also supported by our inelastic neutron scattering results revealing an enhanced low-energy spectral weight along ($H$, $K$, 1/3). The enhanced spectral weight around (0, 0, 1/2) may further originate from competing nested Fermi surface sheets at larger energy transfers (see SI Note 6 for details).   

\section*{Conclusion}

The intricate interaction between the conduction electrons and localized moments in magnetic Kondo metals allow for the exploration of novel frustrated phases, whose interpretation remains challenging. In YbAgGe we observe enhanced dynamic fluctuations of 4$f$ Yb-moments below $T^* \approx$ 20 K. Below $T_{SR}\approx$ 2 K the interaction between the itinerant electrons and local magnetic moments are modified, giving rise to a liquid-like state, which can be partially modeled via a localised Heisenberg Hamiltonian with AFM intraplane couplings. Magnetic long-range order is established below $T_N$ = 0.68 K, suggesting AFM nearest-neighbour interplane couplings and finite further-neighbour terms. Altogether these experimental findings allude the following hierarchy of rather close energy scales $E_{CEF}^{splitting}>E_{KLH}^{exchange} \geq E_{frustrated}^{interactions}$. The case of YbAgGe testifies that effective model Hamiltonians that encode the electronic Fermi surface are required to accurately predict the microscopic properties of frustrated Kondo metals. 

\section*{Methods}

\textbf{Crystal synthesis and macroscopic measurements.} YbAgGe single crystals were grown from Ag- and Ge-rich high-temperature solutions and characterized as described in Ref. \cite{Morosan2004}. Sizable single crystals appropriate for inelastic neutron scattering were obtained through optimization of the synthesis process such as described in Ref. \cite{Larsen2021} and \cite{Canfield2019} (see SI Note 1). Electric resistivity was measured down to $T$ = 1.8 K using a Quantum Design PPMS-14 instrument. We used a standard AC four probe resistance technique ($f$ = 16 Hz, $I$ = 1 -0.3 mA), where Pt leads were attached to a surface-polished YbAgGe single crystal using  Epotek H2OE silver epoxy so that the current was flowing either along the crystallographic $c$-direction or within the $ab$-plane. The magnetization experiments in pulsed magnetic fields up to $\mu_0H$ = 50 T were performed at the Dresden High Magnetic Field Laboratory. We used a compensated pickup-coil
system in a pulse-field magnetometer with a home-built $^4$He cryostat and a pulsed magnet with an inner bore of 20 mm that was powered by a 1.44 MJ capacitor bank\cite{Skourski2011}. 

\textbf{Crystal electric-field measurements.} The crystal electric-field scheme of YbAgGe was probed at $T$ = 3 K on a powdered sample enclosed in an Al can. The data were measured on the time-of-flight spectrometer MARI at ISIS Neutron and Muon Source, UK, using an incident neutron beam of $E_i$ = 49.75 meV for which we found an elastic energy resolution of 3.98(2) meV full-width at half-maximum (FWHM), which is larger than the intrinsic resolution of the spectrometer. The spectrum with a wavevector transfer $|q| < $ 4 \AA$^{-1}$ was fitted to two Gaussian peaks revealing CEF transitions centered at $E$ = 12.0(4) meV. An additional peak emerges at $E$ = 23(1) meV for 6 $< |q| < $ 10 \AA$^{-1}$. The energy spectrum was analyzed using the PyCrystalField software package \cite{scheie2021}. Further information can be found in SI Note 2.

\textbf{Static and dynamic magnetism at zero magnetic field.} The static and dynamic properties at $\mu_0H$ =  0 T and $T$ = 0.25 - 30 K were studied on the Cold Neutron Chopper Spectrometer (CNCS) at the Spallation Neutron Source (SNS), Oak Ridge National Laboratory, USA. A $m$ $\sim$2.5 g crystal oriented with ($H$, $H$, $L$) in the horizontal scattering plane was measured for 20 or 40 s per point over an angular range or 180 or 360$^{\circ}$ with a step size of 0.5$^{\circ}$. We used incident energies $E_i$ = 12, 3.32 and 1.55 meV with  chopper frequencies of 180 Hz on both double disk choppers, resulting in elastic energy resolutions $\Delta E$ = 1.2, 0.175 and 0.04 meV FWHM. The data was reduced with the Mantid software package \cite{mantid2014} and folded using the symmetry operations of $P\bar{6}2m$.

The magnetic field dependence of the dynamic properties were measured on the cold neutron multiplexing spectrometer CAMEA at the Paul Scherrer Institut, Switzerland \cite{CAMEA2023}. The sample with $m$ $\sim$4 g consisted of two single crystals that were coaligned within 1$^{\circ}$ in the horizontal ($H$, $H$, $L$) plane. We used a vertical 11 T magnet with dilution insert enabling measurements down to $T$ = 50 mK. For $\mu_0H$ = 0, 2 and 5.5 T we used an incident energy $E_i$ = 5 meV, providing an elastic line resolution of $\Delta E$ = 0.19 meV FWHM (see also SI Note 6). At $\mu_0H$ = 11 T  we used combination of different incident energies $E_i$ = 3.6, 5, 6.8, 7.8, 8.7 and 9.6 meV with elastic line resolution of $\Delta E$ = 0.11 meV FWHM. The 60$^{\circ}$ wide detector was set at angles to cover a reciprocal space range of $\sim$0.5 $<$ $|q|$ $<$  $\sim$2.3 \AA$^{-1}$ in the horizontal scattering plane optimized to follow the magnetic signal. We used a sample rotation step size of 0.5$^{\circ}$ to cover an angular range of 240$^{\circ}$ for detector settings $|q|$ $<$  $\sim$1.5 \AA$^{-1}$, and 120$^{\circ}$ for $\sim$0.8 $<$ $|q|$ $<$  $\sim$2.3 \AA$^{-1}$. About 25 h of statistics was collected at $\mu_0H$ = 0, 2 and 5.5 T, and 71 h for $\mu_0H$ = 11 T. The data was reduced with the MJOLNIR software package \cite{mjlonir2020}. The background subtraction was performed via a powder average over the ($q$, $E$) range where no signal was present, similar to Refs. \cite{Sala2023,Facheris2022}.

\section*{Acknowledgements}
We thank D. McMorrow and Ch. Rüegg for sharing their earlier results on YbAgGe. We acknowledge fruitful discussions with Ch. Rüegg, P. Orth, N. Gauthier, J. Lass, P. Baral, H. D. Rosales, F. A. G. Albarracín, C. Batista and M. Janoschek. This work received financial support from the Swiss National Science Foundation (Grants No. 200020\_182536, 200021\_200653). This research used resources at the Spallation Neutron Source (proposal number IPTS 3207), Department of Energy (DOE) Office of Science User Facilities operated by Oak Ridge National Laboratory (ORNL). We acknowledge the Paul Scherrer Institut for the allocated beamtime on CAMEA (Proposal No. 20212894). We acknowledge the support of the HLD at HZDR, a member of the European Magnetic Field Laboratory (EMFL). Work done at Ames National Laboratory was supported by the U.S. Department of Energy, Office of Basic Energy Science, Division of Materials Sciences and Engineering. Ames National Laboratory is operated for the U.S. Department of Energy by Iowa State University under Contract No. DE-AC02-07CH11358.

\section*{Author contributions}

The project was conceived and initiated by D.G.M., O.Z. P.C.C., S.L.B., B.F. and Ch.R. The samples were synthesised and characterized by P.C.C. and S.L.B., D.G.M., C.B.L., B.U., X.B., R.B., D.M.P, Y.S., J.T., S.L.B., R.M., P.C.C. and O.Z. prepared and performed the experiments. D.G.M., B.U., S.L.B., R.M., P.C.C. and O.Z. analyzed and interpreted the experimental data. The paper was written by D.G.M., B.U., S.L.B., R.M., P.C.C. and O.Z. with the input from all co-authors.

\section*{Additional Information}
Correspondence and requests for materials should be addressed to D.G.M. or O.Z.
\section*{Competing financial interests}
The authors declare no competing interests.
\section*{Data availability}
The experimental data used in this work can be found using the link \href{https://doi.org/10.5281/zenodo.19219126}{https://doi.org/10.5281/zenodo.19219126}

\section*{Code availability}
The code used to analyze the data presented in this study is available from the authors upon reasonable request.
\bibliographystyle{naturemag}
\bibliography{YbAgGebib}

\begin{thebibliography}{10}
\expandafter\ifx\csname url\endcsname\relax
  \def\url#1{\texttt{#1}}\fi
\expandafter\ifx\csname urlprefix\endcsname\relax\def\urlprefix{URL }\fi
\providecommand{\bibinfo}[2]{#2}
\providecommand{\eprint}[2][]{\url{#2}}

\bibitem{Fert2017}
\bibinfo{author}{Fert, A.}, \bibinfo{author}{Reyren, N.} \&
  \bibinfo{author}{Cros, V.}
\newblock \bibinfo{title}{Magnetic skyrmions: advances in physics and potential
  applications}.
\newblock \emph{\bibinfo{journal}{Nature Reviews Materials}}
  \textbf{\bibinfo{volume}{2}}, \bibinfo{pages}{17031} (\bibinfo{year}{2017}).
\newblock \urlprefix\url{https://doi.org/10.1038/natrevmats.2017.31}.

\bibitem{Mydosh1993}
\bibinfo{author}{Mydosh, J.}
\newblock \emph{\bibinfo{title}{Spin Glasses: An Experimental Introduction}}
  (\bibinfo{publisher}{Taylor \& Francis}, \bibinfo{year}{1993}).
\newblock \urlprefix\url{https://books.google.ch/books?id=NcIYzgEACAAJ}.

\bibitem{Balents2010}
\bibinfo{author}{Balents, L.}
\newblock \bibinfo{title}{Spin liquids in frustrated magnets}.
\newblock \emph{\bibinfo{journal}{Nature}} \textbf{\bibinfo{volume}{464}},
  \bibinfo{pages}{199--208} (\bibinfo{year}{2010}).
\newblock \urlprefix\url{https://doi.org/10.1038/nature08917}.

\bibitem{Starykh2015}
\bibinfo{author}{Starykh, O.~A.}
\newblock \bibinfo{title}{Unusual ordered phases of highly frustrated magnets:
  a review}.
\newblock \emph{\bibinfo{journal}{Reports on Progress in Physics}}
  \textbf{\bibinfo{volume}{78}}, \bibinfo{pages}{052502}
  (\bibinfo{year}{2015}).
\newblock \urlprefix\url{https://dx.doi.org/10.1088/0034-4885/78/5/052502}.

\bibitem{broholm2020}
\bibinfo{author}{Broholm, C.} \emph{et~al.}
\newblock \bibinfo{title}{Quantum spin liquids}.
\newblock \emph{\bibinfo{journal}{Science}} \textbf{\bibinfo{volume}{367}},
  \bibinfo{pages}{eaay0668} (\bibinfo{year}{2020}).

\bibitem{Hayami2021}
\bibinfo{author}{Hayami, S.} \& \bibinfo{author}{Motome, Y.}
\newblock \bibinfo{title}{Topological spin crystals by itinerant frustration}.
\newblock \emph{\bibinfo{journal}{Journal of Physics: Condensed Matter}}
  \textbf{\bibinfo{volume}{33}}, \bibinfo{pages}{443001}
  (\bibinfo{year}{2021}).
\newblock \urlprefix\url{https://dx.doi.org/10.1088/1361-648X/ac1a30}.

\bibitem{Hassanieh2010}
\bibinfo{author}{Al-Hassanieh, K.~A.}, \bibinfo{author}{Yang, Y.-F.},
  \bibinfo{author}{Martin, I.} \& \bibinfo{author}{Batista, C.~D.}
\newblock \bibinfo{title}{Effective low-energy model for $f$-electron
  delocalization}.
\newblock \emph{\bibinfo{journal}{Phys. Rev. Lett.}}
  \textbf{\bibinfo{volume}{105}}, \bibinfo{pages}{086402}
  (\bibinfo{year}{2010}).
\newblock
  \urlprefix\url{https://link.aps.org/doi/10.1103/PhysRevLett.105.086402}.

\bibitem{Stewart2001}
\bibinfo{author}{Stewart, G.~R.}
\newblock \bibinfo{title}{Non-{F}ermi-liquid behavior in $d$- and $f$-electron
  metals}.
\newblock \emph{\bibinfo{journal}{Rev. Mod. Phys.}}
  \textbf{\bibinfo{volume}{73}}, \bibinfo{pages}{797--855}
  (\bibinfo{year}{2001}).
\newblock \urlprefix\url{https://link.aps.org/doi/10.1103/RevModPhys.73.797}.

\bibitem{Ruderman1954}
\bibinfo{author}{Ruderman, M.~A.} \& \bibinfo{author}{Kittel, C.}
\newblock \bibinfo{title}{Indirect exchange coupling of nuclear magnetic
  moments by conduction electrons}.
\newblock \emph{\bibinfo{journal}{Phys. Rev.}} \textbf{\bibinfo{volume}{96}},
  \bibinfo{pages}{99--102} (\bibinfo{year}{1954}).
\newblock \urlprefix\url{https://link.aps.org/doi/10.1103/PhysRev.96.99}.

\bibitem{Kasuya1956}
\bibinfo{author}{Kasuya, T.}
\newblock \bibinfo{title}{{A Theory of Metallic Ferro- and Antiferromagnetism
  on Zener's Model}}.
\newblock \emph{\bibinfo{journal}{Progress of Theoretical Physics}}
  \textbf{\bibinfo{volume}{16}}, \bibinfo{pages}{45--57}
  (\bibinfo{year}{1956}).
\newblock \urlprefix\url{https://doi.org/10.1143/PTP.16.45}.
\newblock
  \eprint{https://academic.oup.com/ptp/article-pdf/16/1/45/5266722/16-1-45.pdf}.

\bibitem{Yosida1957}
\bibinfo{author}{Yosida, K.}
\newblock \bibinfo{title}{Magnetic properties of {C}u-{M}n alloys}.
\newblock \emph{\bibinfo{journal}{Phys. Rev.}} \textbf{\bibinfo{volume}{106}},
  \bibinfo{pages}{893--898} (\bibinfo{year}{1957}).
\newblock \urlprefix\url{https://link.aps.org/doi/10.1103/PhysRev.106.893}.

\bibitem{Lacroix2010}
\bibinfo{author}{Lacroix, C.}
\newblock \bibinfo{title}{Frustrated metallic systems: A review of some
  peculiar behavior}.
\newblock \emph{\bibinfo{journal}{Journal of the Physical Society of Japan}}
  \textbf{\bibinfo{volume}{79}}, \bibinfo{pages}{011008}
  (\bibinfo{year}{2010}).
\newblock \urlprefix\url{https://doi.org/10.1143/JPSJ.79.011008}.
\newblock \eprint{https://doi.org/10.1143/JPSJ.79.011008}.

\bibitem{Stockert2020}
\bibinfo{author}{Stockert, O.} \emph{et~al.}
\newblock \bibinfo{title}{Magnetic frustration in a metallic $fcc$ lattice}.
\newblock \emph{\bibinfo{journal}{Phys. Rev. Res.}}
  \textbf{\bibinfo{volume}{2}}, \bibinfo{pages}{013183} (\bibinfo{year}{2020}).
\newblock
  \urlprefix\url{https://link.aps.org/doi/10.1103/PhysRevResearch.2.013183}.

\bibitem{Kurumaji2019}
\bibinfo{author}{Kurumaji, T.} \emph{et~al.}
\newblock \bibinfo{title}{Skyrmion lattice with a giant topological {H}all
  effect in a frustrated triangular-lattice magnet}.
\newblock \emph{\bibinfo{journal}{Science}} \textbf{\bibinfo{volume}{365}},
  \bibinfo{pages}{914--918} (\bibinfo{year}{2019}).
\newblock
  \urlprefix\url{https://www.science.org/doi/abs/10.1126/science.aau0968}.
\newblock \eprint{https://www.science.org/doi/pdf/10.1126/science.aau0968}.

\bibitem{Gao2016}
\bibinfo{author}{Gao, S.} \emph{et~al.}
\newblock \bibinfo{title}{Spiral spin-liquid and the emergence of a vortex-like
  state in {M}n{S}c$_2${S}$_4$}.
\newblock \emph{\bibinfo{journal}{Nature Physics}}
  \textbf{\bibinfo{volume}{13}}, \bibinfo{pages}{157--161}
  (\bibinfo{year}{2017}).
\newblock \urlprefix\url{https://doi.org/10.1038/nphys3914}.

\bibitem{Budko2004}
\bibinfo{author}{Bud'ko, S.~L.}, \bibinfo{author}{Morosan, E.} \&
  \bibinfo{author}{Canfield, P.~C.}
\newblock \bibinfo{title}{Magnetic field induced non-{F}ermi-liquid behavior in
  {YbAgGe} single crystals}.
\newblock \emph{\bibinfo{journal}{Phys. Rev. B}} \textbf{\bibinfo{volume}{69}},
  \bibinfo{pages}{014415} (\bibinfo{year}{2004}).
\newblock \urlprefix\url{https://link.aps.org/doi/10.1103/PhysRevB.69.014415}.

\bibitem{Kondo1964}
\bibinfo{author}{Kondo, J.}
\newblock \bibinfo{title}{Resistance minimum in dilute magnetic alloys}.
\newblock \emph{\bibinfo{journal}{Progress of Theoretical Physics}}
  \textbf{\bibinfo{volume}{32}}, \bibinfo{pages}{37--49}
  (\bibinfo{year}{1964}).
\newblock \urlprefix\url{https://doi.org/10.1143/PTP.32.37}.
\newblock
  \eprint{https://academic.oup.com/ptp/article-pdf/32/1/37/5193092/32-1-37.pdf}.

\bibitem{hewson1997kondo}
\bibinfo{author}{Hewson, A.~C.}
\newblock \emph{\bibinfo{title}{The {K}ondo problem to heavy fermions}}.
\newblock \bibinfo{number}{2} (\bibinfo{publisher}{Cambridge university press},
  \bibinfo{year}{1997}).

\bibitem{Gibson1996}
\bibinfo{author}{Gibson, B.}, \bibinfo{author}{P{\"o}ttgen, R.},
  \bibinfo{author}{Kremer, R.~K.}, \bibinfo{author}{Simon, A.} \&
  \bibinfo{author}{Ziebeck, K. R.~A.}
\newblock \bibinfo{title}{Ternary germanides {LnAgGe} ({Ln} = {Y, Sm, Gd}-{Lu})
  with ordered {Fe}$_{2}${P}-type structure}.
\newblock \emph{\bibinfo{journal}{J. Alloys Compounds}}
  \textbf{\bibinfo{volume}{239}}, \bibinfo{pages}{34} (\bibinfo{year}{1996}).

\bibitem{Matsumura2004}
\bibinfo{author}{Matsumura, T.} \emph{et~al.}
\newblock \bibinfo{title}{Spin fluctuation and crystal field excitation of
  heavy-fermion compound {YbAgGe} studied by inelastic neutron scattering}.
\newblock \emph{\bibinfo{journal}{Journal of the Physical Society of Japan}}
  \textbf{\bibinfo{volume}{73}}, \bibinfo{pages}{2967--2970}
  (\bibinfo{year}{2004}).
\newblock \urlprefix\url{https://doi.org/10.1143/JPSJ.73.2967}.
\newblock \eprint{https://doi.org/10.1143/JPSJ.73.2967}.

\bibitem{Bonville2007}
\bibinfo{author}{Bonville, P.} \emph{et~al.}
\newblock \bibinfo{title}{Magnetic structures and crystal field in the heavy
  electron materials {YbAgGe} and {YbPtIn}}.
\newblock \emph{\bibinfo{journal}{The European Physical Journal B}}
  \textbf{\bibinfo{volume}{55}}, \bibinfo{pages}{77--84}
  (\bibinfo{year}{2007}).
\newblock \urlprefix\url{https://doi.org/10.1140/epjb/e2007-00042-6}.

\bibitem{scheie2021}
\bibinfo{author}{Scheie, A.}
\newblock \bibinfo{title}{{{\it PyCrystalField}: software for calculation,
  analysis and fitting of crystal electric field Hamiltonians}}.
\newblock \emph{\bibinfo{journal}{Journal of Applied Crystallography}}
  \textbf{\bibinfo{volume}{54}}, \bibinfo{pages}{356--362}
  (\bibinfo{year}{2021}).
\newblock \urlprefix\url{https://doi.org/10.1107/S160057672001554X}.

\bibitem{Larsen2021}
\bibinfo{author}{Larsen, C.~B.} \emph{et~al.}
\newblock \bibinfo{title}{Ubiquity of amplitude-modulated magnetic ordering in
  the {H}-{T} phase diagram of the frustrated non-{F}ermi-liquid {YbAgGe}}.
\newblock \emph{\bibinfo{journal}{Phys. Rev. B}}
  \textbf{\bibinfo{volume}{104}}, \bibinfo{pages}{054424}
  (\bibinfo{year}{2021}).
\newblock \urlprefix\url{https://link.aps.org/doi/10.1103/PhysRevB.104.054424}.

\bibitem{Larsen2023}
\bibinfo{author}{Larsen, C.~B.} \emph{et~al.}
\newblock \bibinfo{title}{Magnetic ground state and perturbations of the
  distorted kagome {I}sing metal {TmAgGe}}.
\newblock \emph{\bibinfo{journal}{Phys. Rev. B}}
  \textbf{\bibinfo{volume}{107}}, \bibinfo{pages}{224419}
  (\bibinfo{year}{2023}).
\newblock \urlprefix\url{https://link.aps.org/doi/10.1103/PhysRevB.107.224419}.

\bibitem{Paddison2018}
\bibinfo{author}{Paddison, J. A.~M.}
\newblock \bibinfo{title}{Ultrafast calculation of diffuse scattering from
  atomistic models}.
\newblock \emph{\bibinfo{journal}{Acta. Cryst. A}}
  \textbf{\bibinfo{volume}{75}}, \bibinfo{pages}{14--24}
  (\bibinfo{year}{2019}).
\newblock \urlprefix\url{https://doi.org/10.1107/S2053273318015632}.

\bibitem{Morosan2004}
\bibinfo{author}{Morosan, E.}, \bibinfo{author}{Bud'ko, S.},
  \bibinfo{author}{Canfield, P.}, \bibinfo{author}{Torikachvili, M.} \&
  \bibinfo{author}{Lacerda, A.}
\newblock \bibinfo{title}{Thermodynamic and transport properties of {RAgGe}
  ({R}={Tb-Lu}) single crystals}.
\newblock \emph{\bibinfo{journal}{Journal of Magnetism and Magnetic Materials}}
  \textbf{\bibinfo{volume}{277}}, \bibinfo{pages}{298--321}
  (\bibinfo{year}{2004}).
\newblock
  \urlprefix\url{https://www.sciencedirect.com/science/article/pii/S0304885303009260}.

\bibitem{Fak2005}
\bibinfo{author}{Fåk, B.} \emph{et~al.}
\newblock \bibinfo{title}{An inelastic neutron scattering study of
  single-crystal heavy-fermion {YbAgGe}}.
\newblock \emph{\bibinfo{journal}{Journal of Physics: Condensed Matter}}
  \textbf{\bibinfo{volume}{17}}, \bibinfo{pages}{301} (\bibinfo{year}{2004}).
\newblock \urlprefix\url{https://dx.doi.org/10.1088/0953-8984/17/2/006}.

\bibitem{Amato1988}
\bibinfo{author}{Rossat-Mignod, J.} \emph{et~al.}
\newblock \bibinfo{title}{Inelastic neutron scattering study of cerium heavy
  fermion compounds}.
\newblock \emph{\bibinfo{journal}{Journal of Magnetism and Magnetic Materials}}
  \textbf{\bibinfo{volume}{76-77}}, \bibinfo{pages}{376--384}
  (\bibinfo{year}{1988}).
\newblock
  \urlprefix\url{https://www.sciencedirect.com/science/article/pii/0304885388904295}.

\bibitem{Kadowaki2004}
\bibinfo{author}{Kadowaki, H.}, \bibinfo{author}{Sato, M.} \&
  \bibinfo{author}{Kawarazaki, S.}
\newblock \bibinfo{title}{Spin fluctuation in heavy fermion
  {CeRu}$_{2}${Si}$_{2}$}.
\newblock \emph{\bibinfo{journal}{Phys. Rev. Lett.}}
  \textbf{\bibinfo{volume}{92}}, \bibinfo{pages}{097204}
  (\bibinfo{year}{2004}).
\newblock
  \urlprefix\url{https://link.aps.org/doi/10.1103/PhysRevLett.92.097204}.

\bibitem{Knafo2004}
\bibinfo{author}{Knafo, W.} \emph{et~al.}
\newblock \bibinfo{title}{Anomalous scaling behavior of the dynamical spin
  susceptibility of ce$_{0.925}$la$_{0.075}$ru$_{2}$si$_{2}$}.
\newblock \emph{\bibinfo{journal}{Phys. Rev. B}} \textbf{\bibinfo{volume}{70}},
  \bibinfo{pages}{174401} (\bibinfo{year}{2004}).
\newblock \urlprefix\url{https://link.aps.org/doi/10.1103/PhysRevB.70.174401}.

\bibitem{Boraley2025}
\bibinfo{author}{Boraley, X.} \emph{et~al.}
\newblock \bibinfo{title}{Microscopic origin of reduced magnetic order in a
  frustrated metal}.
\newblock \emph{\bibinfo{journal}{Phys. Rev. Lett.}}
  \textbf{\bibinfo{volume}{135}}, \bibinfo{pages}{046702}
  (\bibinfo{year}{2025}).
\newblock \urlprefix\url{https://link.aps.org/doi/10.1103/38ds-xjl3}.

\bibitem{Halloran2025}
\bibinfo{author}{Halloran, T.} \emph{et~al.}
\newblock \bibinfo{title}{Connection between $f$-electron correlations and
  magnetic excitations in {UT}e$_{2}$}.
\newblock \emph{\bibinfo{journal}{npj Quantum Materials}}
  \textbf{\bibinfo{volume}{10}}, \bibinfo{pages}{2} (\bibinfo{year}{2025}).
\newblock \urlprefix\url{https://doi.org/10.1038/s41535-024-00720-9}.

\bibitem{Chen2020}
\bibinfo{author}{Chen, X.} \emph{et~al.}
\newblock \bibinfo{title}{Unconventional {H}und metal in a weak itinerant
  ferromagnet}.
\newblock \emph{\bibinfo{journal}{Nature Communications}}
  \textbf{\bibinfo{volume}{11}}, \bibinfo{pages}{3076} (\bibinfo{year}{2020}).
\newblock \urlprefix\url{https://doi.org/10.1038/s41467-020-16868-4}.

\bibitem{ROSSATMIGNOD1988376}
\bibinfo{author}{Rossat-Mignod, J.} \emph{et~al.}
\newblock \bibinfo{title}{Inelastic neutron scattering study of cerium heavy
  fermion compounds}.
\newblock \emph{\bibinfo{journal}{Journal of Magnetism and Magnetic Materials}}
  \textbf{\bibinfo{volume}{76-77}}, \bibinfo{pages}{376--384}
  (\bibinfo{year}{1988}).

\bibitem{Riberolles2024}
\bibinfo{author}{Riberolles, S. X.~M.} \emph{et~al.}
\newblock \bibinfo{title}{Chiral and flat-band magnetic quasiparticles in
  ferromagnetic and metallic kagome layers}.
\newblock \emph{\bibinfo{journal}{Nature Communications}}
  \textbf{\bibinfo{volume}{15}}, \bibinfo{pages}{1592} (\bibinfo{year}{2024}).
\newblock \urlprefix\url{https://doi.org/10.1038/s41467-024-45841-8}.

\bibitem{Riberolles2023}
\bibinfo{author}{Riberolles, S. X.~M.} \emph{et~al.}
\newblock \bibinfo{title}{Orbital character of the spin-reorientation
  transition in {T}b{M}n$_{6}${S}s$_{6}$}.
\newblock \emph{\bibinfo{journal}{Nature Communications}}
  \textbf{\bibinfo{volume}{14}}, \bibinfo{pages}{2658} (\bibinfo{year}{2023}).

\bibitem{Riberolles2022}
\bibinfo{author}{Riberolles, S. X.~M.} \emph{et~al.}
\newblock \bibinfo{title}{Low-temperature competing magnetic energy scales in
  the topological ferrimagnet {T}b{M}n$_{6}${S}s$_{6}$}.
\newblock \emph{\bibinfo{journal}{Phys. Rev. X}} \textbf{\bibinfo{volume}{12}},
  \bibinfo{pages}{021043} (\bibinfo{year}{2022}).
\newblock \urlprefix\url{https://link.aps.org/doi/10.1103/PhysRevX.12.021043}.

\bibitem{Goremychkin2018}
\bibinfo{author}{Goremychkin, E.~A.} \emph{et~al.}
\newblock \bibinfo{title}{Coherent band excitations in {C}e{P}d$_3$: A
  comparison of neutron scattering and $ab$ $initio$ theory}.
\newblock \emph{\bibinfo{journal}{Science}} \textbf{\bibinfo{volume}{359}},
  \bibinfo{pages}{186--191} (\bibinfo{year}{2018}).
\newblock
  \urlprefix\url{https://www.science.org/doi/abs/10.1126/science.aan0593}.
\newblock \eprint{https://www.science.org/doi/pdf/10.1126/science.aan0593}.

\bibitem{Simeth2023}
\bibinfo{author}{Simeth, W.} \emph{et~al.}
\newblock \bibinfo{title}{A microscopic {K}ondo lattice model for the heavy
  fermion antiferromagnet {C}e{I}n$_3$}.
\newblock \emph{\bibinfo{journal}{Nature Communications}}
  \textbf{\bibinfo{volume}{14}}, \bibinfo{pages}{8239} (\bibinfo{year}{2023}).
\newblock \urlprefix\url{https://doi.org/10.1038/s41467-023-43947-z}.

\bibitem{Ghioldi2024}
\bibinfo{author}{Ghioldi, E.~A.} \emph{et~al.}
\newblock \bibinfo{title}{Derivation of low-energy {H}amiltonians for
  heavy-fermion materials}.
\newblock \emph{\bibinfo{journal}{Phys. Rev. B}}
  \textbf{\bibinfo{volume}{110}}, \bibinfo{pages}{195123}
  (\bibinfo{year}{2024}).
\newblock \urlprefix\url{https://link.aps.org/doi/10.1103/PhysRevB.110.195123}.

\bibitem{Kan2020}
\bibinfo{author}{Zhao, K.} \emph{et~al.}
\newblock \bibinfo{title}{Realization of the kagome spin ice state in a
  frustrated intermetallic compound}.
\newblock \emph{\bibinfo{journal}{Science}} \textbf{\bibinfo{volume}{367}},
  \bibinfo{pages}{1218--1223} (\bibinfo{year}{2020}).
\newblock
  \urlprefix\url{https://www.science.org/doi/abs/10.1126/science.aaw1666}.
\newblock \eprint{https://www.science.org/doi/pdf/10.1126/science.aaw1666}.

\bibitem{Bhandari2024}
\bibinfo{author}{Bhandari, H.} \emph{et~al.}
\newblock \bibinfo{title}{Tunable topological transitions in the frustrated
  magnet {HoAgGe}} (\bibinfo{year}{2024}).
\newblock \urlprefix\url{https://arxiv.org/abs/2410.11636}.
\newblock \eprint{2410.11636}.

\bibitem{Schwertfeger2025}
\bibinfo{author}{Schwertfeger, G.~F.}, \bibinfo{author}{Chang, P.-H.},
  \bibinfo{author}{Nikoli\ifmmode~\acute{c}\else \'{c}\fi{}, P.} \&
  \bibinfo{author}{Mazin, I.~I.}
\newblock \bibinfo{title}{Modeling of a twisted kagome hoagge spin ice using
  reduced configuration space search and density functional theory}.
\newblock \emph{\bibinfo{journal}{Phys. Rev. B}}
  \textbf{\bibinfo{volume}{112}}, \bibinfo{pages}{214411}
  (\bibinfo{year}{2025}).
\newblock \urlprefix\url{https://link.aps.org/doi/10.1103/p8bq-9623}.

\bibitem{Canfield2019}
\bibinfo{author}{Canfield, P.~C.}
\newblock \bibinfo{title}{New materials physics}.
\newblock \emph{\bibinfo{journal}{Reports on Progress in Physics}}
  \textbf{\bibinfo{volume}{83}}, \bibinfo{pages}{016501}
  (\bibinfo{year}{2019}).
\newblock \urlprefix\url{https://dx.doi.org/10.1088/1361-6633/ab514b}.

\bibitem{Skourski2011}
\bibinfo{author}{Skourski, Y.}, \bibinfo{author}{Kuz'min, M.~D.},
  \bibinfo{author}{Skokov, K.~P.}, \bibinfo{author}{Andreev, A.~V.} \&
  \bibinfo{author}{Wosnitza, J.}
\newblock \bibinfo{title}{High-field magnetization of {Ho}$_2${Fe}$_{17}$}.
\newblock \emph{\bibinfo{journal}{Phys. Rev. B}} \textbf{\bibinfo{volume}{83}},
  \bibinfo{pages}{214420} (\bibinfo{year}{2011}).
\newblock \urlprefix\url{https://link.aps.org/doi/10.1103/PhysRevB.83.214420}.

\bibitem{mantid2014}
\bibinfo{author}{Arnold, O.} \emph{et~al.}
\newblock \bibinfo{title}{Mantid-data analysis and visualization package for
  neutron scattering and $\mu${SR} experiments}.
\newblock \emph{\bibinfo{journal}{Nuclear Instruments and Methods in Physics
  Research. Section A: Accelerators, Spectrometers, Detectors and Associated
  Equipment}} \textbf{\bibinfo{volume}{764}}, \bibinfo{pages}{156--166}
  (\bibinfo{year}{2014}).
\newblock
  \urlprefix\url{https://www.sciencedirect.com/science/article/pii/S0168900214008729}.

\bibitem{CAMEA2023}
\bibinfo{author}{Lass, J.} \emph{et~al.}
\newblock \bibinfo{title}{Commissioning of the novel {C}ontinuous {A}ngle
  {M}ulti-{E}nergy {A}nalysis spectrometer at the {P}aul {S}cherrer
  {I}nstitut}.
\newblock \emph{\bibinfo{journal}{Rev. Sci. Instrum.}}
  \textbf{\bibinfo{volume}{94}}, \bibinfo{pages}{023302}
  (\bibinfo{year}{2023}).
\newblock \urlprefix\url{https://doi.org/10.1063/5.0128226}.

\bibitem{mjlonir2020}
\bibinfo{author}{Lass, J.}, \bibinfo{author}{Jacobsen, H.},
  \bibinfo{author}{Mazzone, D.~G.} \& \bibinfo{author}{Lefmann, K.}
\newblock \bibinfo{title}{Mjolnir: A software package for multiplexing neutron
  spectrometers}.
\newblock \emph{\bibinfo{journal}{SoftwareX}} \textbf{\bibinfo{volume}{12}},
  \bibinfo{pages}{100600} (\bibinfo{year}{2020}).
\newblock
  \urlprefix\url{https://www.sciencedirect.com/science/article/pii/S2352711020303137}.

\bibitem{Sala2023}
\bibinfo{author}{Sala, G.} \emph{et~al.}
\newblock \bibinfo{title}{Field-tuned quantum renormalization of spin dynamics
  in the honeycomb lattice {H}eisenberg antiferromagnet {Y}b{C}l$_3$}.
\newblock \emph{\bibinfo{journal}{Communications Physics}}
  \textbf{\bibinfo{volume}{6}}, \bibinfo{pages}{234} (\bibinfo{year}{2023}).
\newblock \urlprefix\url{https://doi.org/10.1038/s42005-023-01333-7}.

\bibitem{Facheris2022}
\bibinfo{author}{Facheris, L.} \emph{et~al.}
\newblock \bibinfo{title}{Spin density wave versus fractional magnetization
  plateau in a triangular antiferromagnet}.
\newblock \emph{\bibinfo{journal}{Phys. Rev. Lett.}}
  \textbf{\bibinfo{volume}{129}}, \bibinfo{pages}{087201}
  (\bibinfo{year}{2022}).
\newblock
  \urlprefix\url{https://link.aps.org/doi/10.1103/PhysRevLett.129.087201}.

\bibitem{Schmiedeshoff2011}
\bibinfo{author}{Schmiedeshoff, G.~M.} \emph{et~al.}
\newblock \bibinfo{title}{Multiple regions of quantum criticality in {YbAgGe}}.
\newblock \emph{\bibinfo{journal}{Phys. Rev. B}} \textbf{\bibinfo{volume}{83}},
  \bibinfo{pages}{180408} (\bibinfo{year}{2011}).
\newblock \urlprefix\url{https://link.aps.org/doi/10.1103/PhysRevB.83.180408}.

\bibitem{Toth2015}
\bibinfo{author}{Toth, L.~B., S.}
\newblock \bibinfo{title}{Linear spin wave theory for single-q incommensurate
  magnetic structures}.
\newblock \emph{\bibinfo{journal}{J. Phys.: Condens. Matter}}
  \textbf{\bibinfo{volume}{27}}, \bibinfo{pages}{166002}
  (\bibinfo{year}{2015}).
\newblock \urlprefix\url{https://doi.org/10.1088/0953-8984/27/16/166002}.

\bibitem{Goodenough1963}
\bibinfo{author}{Weiss, A.}
\newblock \bibinfo{title}{John b. goodenough: Magnetism and the chemical bond.
  interscience publishers. new york, london 1963. 393 seiten, 89 abbildungen.
  preis: Dm 95 s.}
\newblock \emph{\bibinfo{journal}{Berichte der Bunsengesellschaft f{\"u}r
  physikalische Chemie}} \textbf{\bibinfo{volume}{68}},
  \bibinfo{pages}{996--996} (\bibinfo{year}{1964}).

\end{thebibliography}

\newpage

\newcommand{\beginsupplement}{
        \setcounter{table}{0}
        \renewcommand{\thetable}{S\arabic{table}}
        \setcounter{figure}{0}
        \renewcommand{\figurename}{\textbf{Supplemental Figure}}}

\graphicspath{{./}{Supp. Mat./}}

\beginsupplement
\title{Supplementary information: Intricacies of Frustrated Magnetism in the Kondo Metal YbAgGe}
\author{D.~G.~Mazzone}
\email{daniel.mazzone@psi.ch}
\affiliation{PSI Center for Neutron and Muon Sciences, 5232 Villigen PSI, Switzerland} 

\author{C.~B.~Larsen}
\affiliation{PSI Center for Neutron and Muon Sciences, 5232 Villigen PSI, Switzerland} 

\author{B. Ueland}
\affiliation{Department of Physics and Astronomy, Iowa State University, Ames, Iowa 50011, USA}
\affiliation{Ames National Laboratory, Iowa State University, Ames, Iowa 50011, USA}

\author{X.~Boraley}
\affiliation{PSI Center for Neutron and Muon Sciences, 5232 Villigen PSI, Switzerland} 

\author{D. M. Pajerowski}
\affiliation{Neutron Scattering Division, Oak Ridge National Laboratory, Oak Ridge, Tennessee 37831, USA}

\author{Y. Skourski}
\affiliation{Dresden High Magnetic Field Laboratory (HLD-EMFL),
Helmholtz-Zentrum Dresden-Rossendorf, 01328 Dresden, Germany}

\author{J. Taylor}
\affiliation{ISIS Facility, STFC Rutherford Appleton Laboratory,
Harwell Science and Innovation Campus, Oxfordshire OX11 0QX, United Kingdom}
\affiliation{Neutron Scattering Division, Oak Ridge National Laboratory, Oak Ridge, Tennessee 37831, USA}

\author{B.~F\aa k}
\affiliation{Institut Laue-Langevin, 71 Avenue des Martyrs, CS20156, 38042 Grenoble C\'edex 9, France}


\author{S. L. Bud’ko}
\affiliation{Department of Physics and Astronomy, Iowa State University, Ames, Iowa 50011, USA}
\affiliation{Ames National Laboratory, Iowa State University, Ames, Iowa 50011, USA}

\author{R. McQueeney}
\affiliation{Department of Physics and Astronomy, Iowa State University, Ames, Iowa 50011, USA}
\affiliation{Ames National Laboratory, Iowa State University, Ames, Iowa 50011, USA}

\author{P. C. Canfield}
\affiliation{Department of Physics and Astronomy, Iowa State University, Ames, Iowa 50011, USA}
\affiliation{Ames National Laboratory, Iowa State University, Ames, Iowa 50011, USA}

\author{O.~Zaharko}
\email{oksana.zaharko@psi.ch}
\affiliation{PSI Center for Neutron and Muon Sciences, 5232 Villigen PSI, Switzerland} 

\maketitle
\onecolumngrid{
\section*{\texorpdfstring{S\MakeLowercase{upplementary} N\MakeLowercase{ote 1:}}{Supplementary Note}  Single crystal growth and magnetic phase diagram}

YbAgGe single crystals were grown from a  Ag- and Ge-rich high-temperature ternary solution \cite{Canfield2019} and characterized based on the procedure described in Ref. \cite{Morosan2004}. In order to grow a significant number of large single crystals of YbAgGe, the growth was optimized in several ways. Roughly 10 g of high purity Yb, Ag, and Ge were placed in a 5 ml Al$_2$O$_3$ crucible in a molar ratio of Yb0:14Ag0:645Ge0:215. The Yb with 99.99+\% purity was obtained from the Ames Laboratory Materials Preparation Center. A second 5 ml Al$_2$O$_3$ crucible, filled with silica wool, was placed on top of the growth crucible to act as a filter and catch for excess liquid \cite{Morosan2004}. The growth and catch crucibles were sealed into an evacuated amorphous silica tube and placed in a resistively heated box furnace.  In order to partially control nucleation, the growth ampoules were placed close to the inner edges of the furnace door.  We found that larger crystals often grew on the side of the crucible facing this colder spot in the furnace.  The furnace was heated to 1100 $^\circ$C over 4 hours, and heated further to 1190 $^\circ$C over an additional hour. The furnace was then cooled to 1090 $^\circ$C over 5 hours and then slow cooled to 840 $^\circ$C over 200 hours. The growth ampoules were removed from the furnace at 840 $^\circ$C and placed in a centrifuge \cite{Canfield2019} for separation of the excess liquid from the YbAgGe single crystals.  Depending on nucleation, single crystals as large as 1 - 2 grams could be obtained (see inset of Supplementary (Suppl.) Fig. \ref{fig:phasediag}).

\begin{figure*}[tbh]
\centering
\includegraphics[width={0.5\textwidth}]{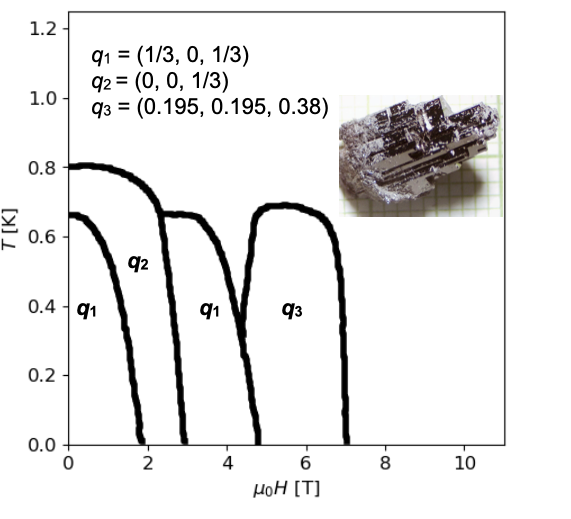}\\
\caption{\label{fig:phasediag}%
\textbf{Schematic $HT$-phase diagram for $H||[1, -1, 0]$.} Schematic magnetic phase diagram of magnetic long-range order in YbAgGe for $H||[1, -1, 0]$ with corresponding magnetic wavevectors $q_{1-3}$. The diagram has been reproduced from Ref. \cite{Larsen2021}. Inset: Picture of a YbAgGe single crystal.}
\end{figure*}

Macroscopic heat capacity and electrical resistivity measurements have found evidence for magnetic correlations below $T_{SR}$ $\approx$ 2 K at zero magnetic field \cite{Budko2004,Morosan2004}. These correlations are signified by a broad response as a function of temperature which is maximal at 0.8 K, and has been used in the past to identify the phase boundary of these correlations in $HT$-phase diagrams. In the main manuscript we show that the temperature region $T_N$ $<$ $T_{SR}$ is associated with short-range order with the propagation vector (0,~0,~{\third}) that is correlated mainly along the $c$-axis. Antiferromagnetic (AFM) long-range order in YbAgGe is established below $T_N$ = 0.68 K with a propagation vector $q$ = ({\third},~0,~{\third}) \cite{Larsen2021}. A magnetic field applied in the Kagome plane triggers a series of metamagnetic transitions to other magnetic states with different periodicity as shown in Suppl. Fig. \ref{fig:phasediag}. Above  $\mu_0H\approx$ 7 T ($H\parallel$ [1, -1, 0]) these modulations vanish, so that eventually only an induced ferromagnetic (FM) component remains. We mention that previous macroscopic studies revealed additional correlated signatures in the $HT$-phase diagram of YbAgGe \cite{Schmiedeshoff2011}.

\section*{\texorpdfstring{S\MakeLowercase{upplementary} N\MakeLowercase{ote 2:}}{Supplementary Note} Crystal electric-field refinement}

The energy spectrum of a polycrystalline YbAgGe sample measured on the MARI spectrometer at ISIS with $E_i$ = 49.75 meV and $T$ = 3 K revealed nondispersing excitations at $E$ = 12.0(4), 23(1) and 36(1) meV.
The first feature reveals intensity prevailing below the wavevector transfers $|q| < $ 4 \AA$^{-1}$, which we attributed to a crystal electric-field (CEF) transition. Fits of this transition against the CEF Hamiltonian $H_{CEF}$ = $\sum_{n,m}B_n^mO_n^m$, with  $O_n^m$ being Stevens operators and $B_n^m$ CEF parameters by PyCrystalField python package  \cite{scheie2021} using the z-direction as the local quantization axis reveal $B_2^0$ $\approx$ 0.137 meV, $B_2^2$ $\approx$ -1.27 meV ($\chi^2$ = 0.17) and negligible higher order CEF parameters. The best fit to the data and resulting set of eignevectors and eigenvalues are shown in Suppl. Fig. \ref{fig:CEF_ref}a and Tabl. \ref{TBL:YbAgGeCEF}, respectively, establishing a ground state doublet $\ket{\Psi^{\pm}}$ = -0.189$\ket{\mp7/2}$ - 0.385$\ket{\pm5/2}$ - 0.578$\ket{\mp3/2}$ - 0.695$\ket{\pm1/2}$ with a localized moment of $\mu_{CEF}$ = 3.98$\mu_B$ and significant axial anisotropy, $g$-tensor = (7.958, 0.0247, 0.0303). Attempts to include the second and third peaks with intensity dominating at 6 $< |q| < $ 10 \AA$^{-1}$ destabilized the refinement and worsens the fit of the $E$ = 12.0(4) meV transition. Along with the $|q|$ dependence of the excitation spectrum this suggests a phononic origin of the $E$ = 23(1) and 36(1) meV excitations (see Suppl. Fig. \ref{fig:CEF_ref}b). The calculated inverse susceptibility $\chi^{-1}$ and magnetization $M$ are shown in Suppl. Figs. \ref{fig:CEF_ref}c and \ref{fig:magnetization}, respectively. In agreement with previous reports \cite{Matsumura2004} the predicted macroscopic quantities reveal some deviations from the experimental results (see Ref. \cite{Matsumura2004}), which have been attributed to itinerant effects in the material and large fluctuations along the $c$-axis.

\begin{figure*}[tbh]
\includegraphics[width=1\textwidth]{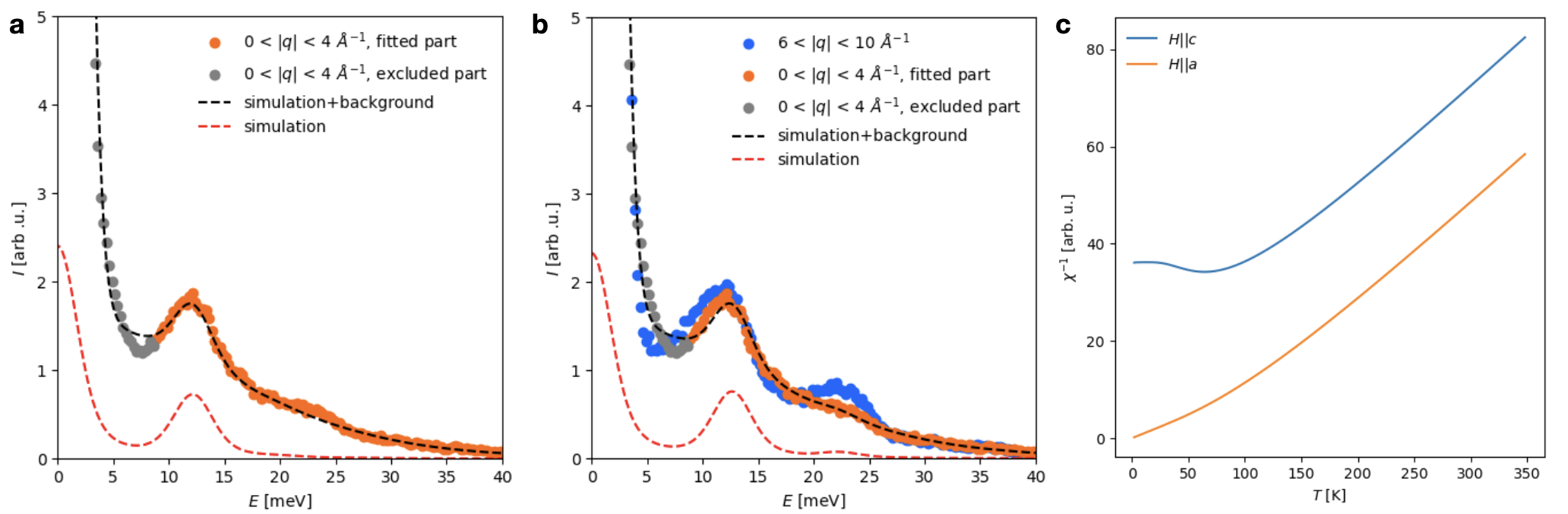}
\caption{\label{fig:CEF_ref}%
\textbf{Crystal electric-field refinement.} \textbf{a} Neutron scattering spectrum obtained on MARI integrated over a wavevector transfer $|q| < $ 4 \AA$^{-1}$ in gray and orange. The orange data points were used to determine the crystal electric-field (CEF) parameters of the material. The black dashed line shows the best fit to the experimental data ($\chi^2$ = 0.17), considering solely the lowest nondispersing excitation at $E$ = 12.0(4) meV (see also Tabl. \ref{TBL:YbAgGeCEF}). The background consists of two Gaussians centered at zero energy transfer, representing the elastic line and monotonically decreasing scattering contribution at increasing energy transfers. The red dashed line shows the predicted excitation spectrum without the background model. \textbf{b} The inclusion of the peaks at $E$ = 23(1) and 36(1) meV with intensity dominating at 6 $< |q| < $ 10 \AA$^{-1}$, yields a  destabilised fit ($\chi^2$ = 58) and worsens the refinement of the $E$ = 12.0(4) meV transition (see black and red dashed lines). \textbf{c} Calculated inverse susceptibility as function of temperature (see Ref. \cite{Matsumura2004} for comparison with earlier refinements and with experimental results).}
\end{figure*}

\begin{table*}
\caption{Eigenvectors and eigenvalues of the CEF scheme of the Yb$^{3+}$ ion in YbAgGe.}
\begin{ruledtabular}
\begin{tabular}{c|cccccccc}
E (meV) &$| -\frac{7}{2}\rangle$ & $| -\frac{5}{2}\rangle$ & $| -\frac{3}{2}\rangle$ & $| -\frac{1}{2}\rangle$ & $| \frac{1}{2}\rangle$ & $| \frac{3}{2}\rangle$ & $| \frac{5}{2}\rangle$ & $| \frac{7}{2}\rangle$ \tabularnewline
 \hline 
0.000 & -0.189 & 0.0 & -0.578 & 0.0 & -0.695 & 0.0 & -0.385 & 0.0 \tabularnewline
0.000 & 0.0 & -0.385 & 0.0 & -0.695 & 0.0 & -0.578 & 0.0 & -0.189 \tabularnewline
12.165 & -0.535 & 0.0 & -0.521 & 0.0 & 0.234 & 0.0 & 0.622 & 0.0 \tabularnewline
12.165 & 0.0 & -0.622 & 0.0 & -0.234 & 0.0 & 0.521 & 0.0 & 0.535 \tabularnewline
19.864 & 0.0 & 0.533 & 0.0 & -0.281 & 0.0 & -0.263 & 0.0 & 0.753 \tabularnewline
19.864 & -0.753 & 0.0 & 0.263 & 0.0 & 0.281 & 0.0 & -0.533 & 0.0 \tabularnewline
27.805 & 0.333 & 0.0 & -0.57 & 0.0 & 0.619 & 0.0 & -0.424 & 0.0 \tabularnewline
27.805 & 0.0 & 0.424 & 0.0 & -0.619 & 0.0 & 0.57 & 0.0 & -0.333 \tabularnewline
\end{tabular}\end{ruledtabular}
\label{TBL:YbAgGeCEF}
\end{table*}

In a localised moment picture the ground state doublet $\ket{\Psi^{\pm}}$ is split by magnetic long-range order. Thus, the dipole allowed transitions between the ground state wavefunctions give an estimation of the neutron cross-section expected in the ordered state. For YbAgGe we find $\bra{\Psi^{+}}J_x\ket{\Psi^{-}}$ = 3.48, $\bra{\Psi^{+}}J_y\ket{\Psi^{-}}$ = 0.0108 and $\bra{\Psi^{+}}J_z\ket{\Psi^{-}}$ = 0. The strongest component indicates the preferred ordered moment direction which is the local x-axis of the three magnetic Yb sites, $i.e.$ (1, 0, 0) for Yb1, (0, 1, 0) for Yb2 and (1, 1, 0) for Yb3 (see also Ref. \cite{Larsen2021}). The other components give a cross-section estimation for the magnetic fluctuations perpendicular to the ordered moment. Their small value indicate very weak spin-wave excitations that are expected to be hardly observable in an inelastic neutron scattering experiment.

\section*{\texorpdfstring{S\MakeLowercase{upplementary} N\MakeLowercase{ote 3:}}{Supplementary Note} Insight into exchange couplings} 

Despite various contributing ingredients in frustrated metals, it is often possible to determine the main couplings through a localised Heisenberg exchange Hamiltonian. For YbAgGe we used the effective Hamiltonian  
\begin{equation}
        H=\sum_{ij} J_{ij}{\bf S}_{i}\cdot{\bf S}_{j}+
        \sum_{ij} J_{cij}{\bf S}_{i}\cdot{\bf S}_{j}+
        A\sum_{i} {\bf S}_{i}^2.
        \label{eq:Hamil}
\end{equation}

The first term corresponds to the exchange couplings $J_{ij}$ between spins $\bf S$ on sites $i$ and $j$ of the distorted Kagome plane. The second term pertains the exchanges with components along the $c$-axis, $J_{cij}$, and the last term accounts for the magnetic anisotropy $A$. 

Our experimental findings and CEF considerations suggest that the energy scale of the magnetic anisotropy and exchange couplings are of the same order in YbAgGe. Thus, the anisotropy does not dominate the selection of the ordering vector. We fixed the parameter to $A$ = -1. Another important experimental fact is that the $q_z$ =  {\third} component of the ordering vector is remarkably persistent in all reported ordered magnetic phases up to $\mu_0H$ = 7 T \cite{Larsen2021}. This suggests that exchange interactions extending along the $c$-axis are strong, competing among themselves and with the intraplane couplings.

We minimized the Hamiltonian of Eq.\eqref{eq:Hamil} using the SpinW software\cite{Toth2015}, restricting the $J$-parameter space to the nearest neighbours presented in Suppl. Fig.~\ref{fig:Exchanges} and scanned the exchanges within $[$-1..1$]$ intervals using a grid step of 0.05. We recorded the minimal energy solutions with propagation vectors corresponding to the experimentally observed vectors $q_1$=({\third},~0,~{\third}), its arms and $q$ = (0, 0, {\third}). 
Firstly, we considered only exchanges up to the second next-nearest neighbours in- and out- the plane, presented in Suppl.  Fig.~\ref{fig:Exchanges}a. The solutions with propagation vectors $q_z$ = {\third} spread over all considered intervals of $J_{1}$, $J_{2}$ and $J_{c1}$, but $J_{c2}$ was confined to $[$-0.5..0.5$]$ region (Suppl. Fig.~\ref{fig:Exchanges}c). With this set of couplings the propagation vector $q_1$=({\third},~0,~{\third}) was not found.
Then we included the equidistant but distinct $J_{3a}$ and $J_{3b}$ in-plane couplings (Suppl. Fig.~\ref{fig:Exchanges}b) into the simulations. The difference between $J_{3a}$ and $J_{3b}$ becomes apparent when inspecting the hexagon encircled by a dashed black line around the $\bar{6}$-axis (cell origin) in Suppl. Fig.~\ref{fig:Exchanges}b. The cyan colored $J_{3a}$ bonds pass though the middle of the hexagon, while the pink $J_{3b}$ bonds span its periphery.
As the $J$-parameter space expanded significantly, we used a cruder grid step of 0.3. The solutions with $q_1$=({\third},~0,~{\third}) were found for finite $J_{3a}$ and $J_{3b}$ values (both positive and negative). 
Interestingly, these equidistant, but different couplings were decisive already in selecting the magnetic field-induced states in TmAgGe\cite{Larsen2023}. 
For YbAgGe the stabilizing the $q_1$=({\third},~0,~{\third}) propagation vector
requires $J_{c1}$ and $J_{c2}$ to be positive, that is AFM (Suppl. Fig.~\ref{fig:Exchanges}d).
\begin{figure*}[tbh]
\includegraphics[width=1\textwidth]{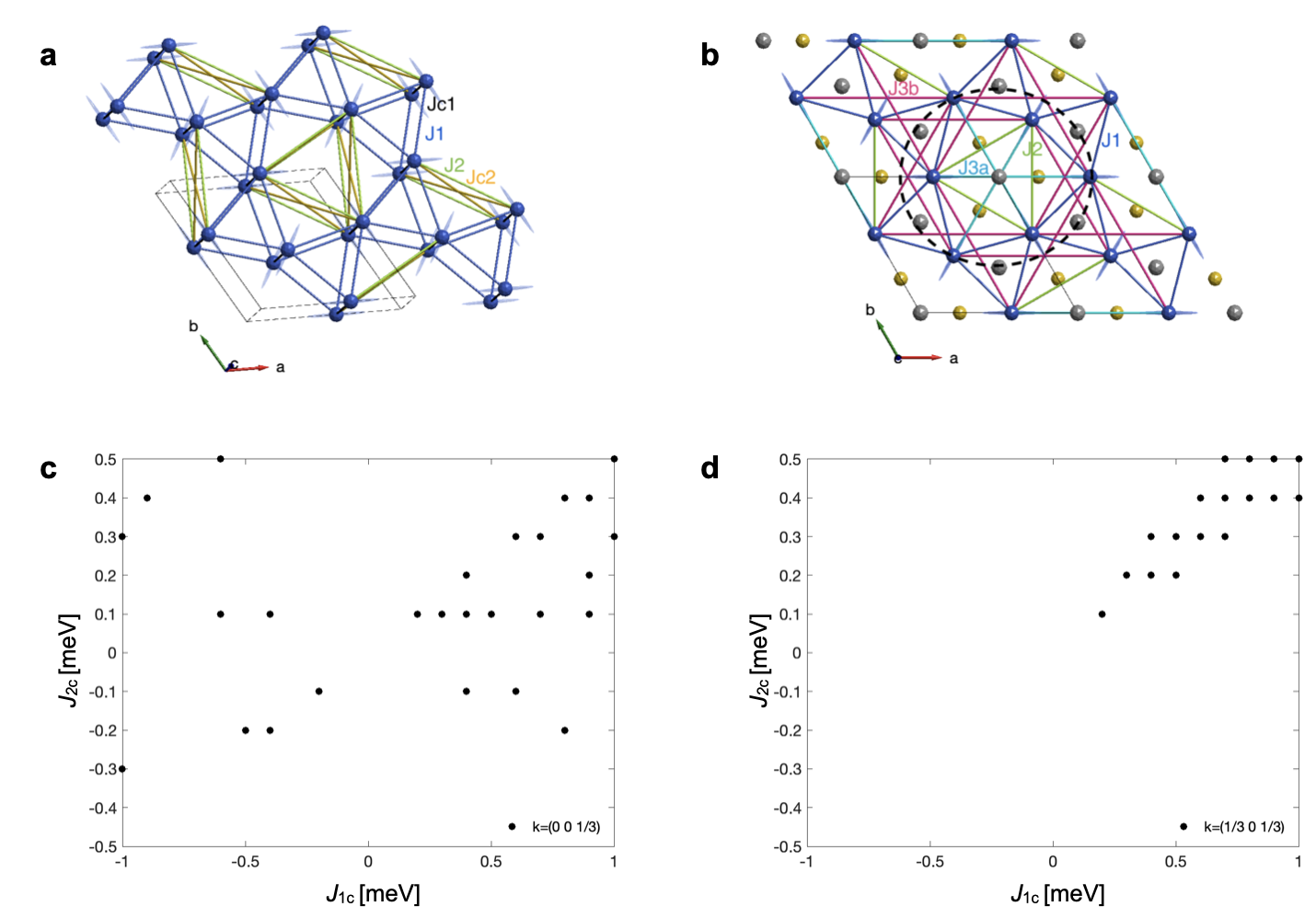}
\caption{\label{fig:Exchanges}%
\textbf{Exchange couplings and results of local Hamiltonian minimization.} \textbf{a} Nearest neighbour exchange couplings $J_1$ (blue), $J_2$ (green) in-plane and  $J_{c1}$ (black), $J_{c2}$ (gold) out-of-plane alongside with the single-ion anisotropy (blue ellipses). \textbf{b} in-plane  $J_1$, $J_2$, $J_{3a}$ (cyan) and $J_{3b}$ (pink) couplings. Yb, Ge and Ag atoms are colored in blue, gray and gold, respectively. The dashed black circle spans over a hexagon of edge sharing Yb-triangles, the elemental building block of the geometrical Yb$^{3+}$ building block. \textbf{c} Distribution of the lowest energy solutions with magnetic ordering vectors $q$= (0,~0,~{\third}) and  \textbf{d} $q_1$= ({\third},~0,~{\third}) as a function of $J_{c1}$ and $J_{c2}$. The other couplings,  $J_1$, $J_2$,  $J_{c1}$ vary within $[$-1 .. 1$]$ intervals, $J_{c2}$ within $[$-0.5 .. 0.5$]$.}
\end{figure*}

\section*{\texorpdfstring{S\MakeLowercase{upplementary} N\MakeLowercase{ote 4:}}{Supplementary Note} High Field Magnetization and comparison to calculations}

The localization degree of the Yb moments was examined through magnetization experiments in pulsed magnetic fields up to $\mu_0H$ = 50 T at the Dresden High Magnetic Field Laboratory, which were normalised to laboratory measurements up to $\mu_0H$ = 14 T reported in Ref.  \cite{Morosan2004}. Supplementary Figure \ref{fig:magnetization} depicts the experimental results measured at $T$ = 0.53 K on a single crystal for $H||ab$ (red line) and $H||c$ (orange line), respectively. Overplotted in green and blue are the calculated magnetization curves expected from the CEF scheme. We assign the deviations at low magnetic fields to contributions of the magnetic exchange interactions that were not considered in the calculations. For $H||ab$ we find qualitative agreement in the overall field dependence and quantitative agreement at large magnetic fields. In contrast, qualitative deviations occur for $H||c$ above $\mu_0H\approx$ 30 T. For $H||c$ we find that the crystal was misaligned by $\sim$12$^\circ$ away from the $c$-axis due to manual alignment of the small crystal in the $^3$He insert. This misalignment is not relevant for the conclusion that the local moment model cannot account for the experimentally observed magnetization curve. 
\begin{figure*}[tbh]
\centering
\includegraphics[width=0.8\textwidth]{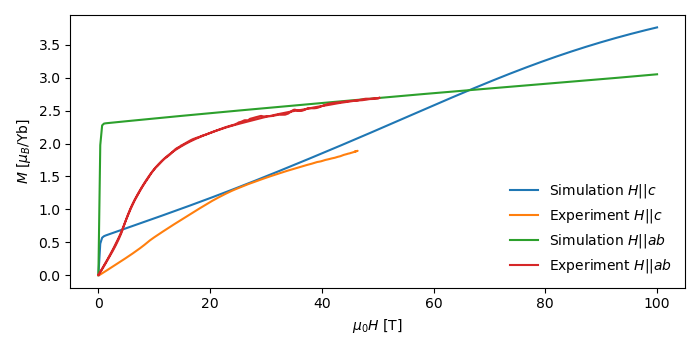}\\
\caption{\label{fig:magnetization}%
\textbf{High-field magnetization.} Field dependent magnetization of YbAgGe for $H||ab$ and $H||c$. The red and orange curves are measurements at $T$= 0.53 K. In green and blue we show the field dependence expected from the crystal electric-field scheme.}
\end{figure*}

\section*{\texorpdfstring{S\MakeLowercase{upplementary} N\MakeLowercase{ote 5:}}{Supplementary Note} Dynamic magnetism in YbAgGe at zero magnetic field}

\begin{figure*}[tbh]
\centering
\includegraphics[width=0.8\textwidth]{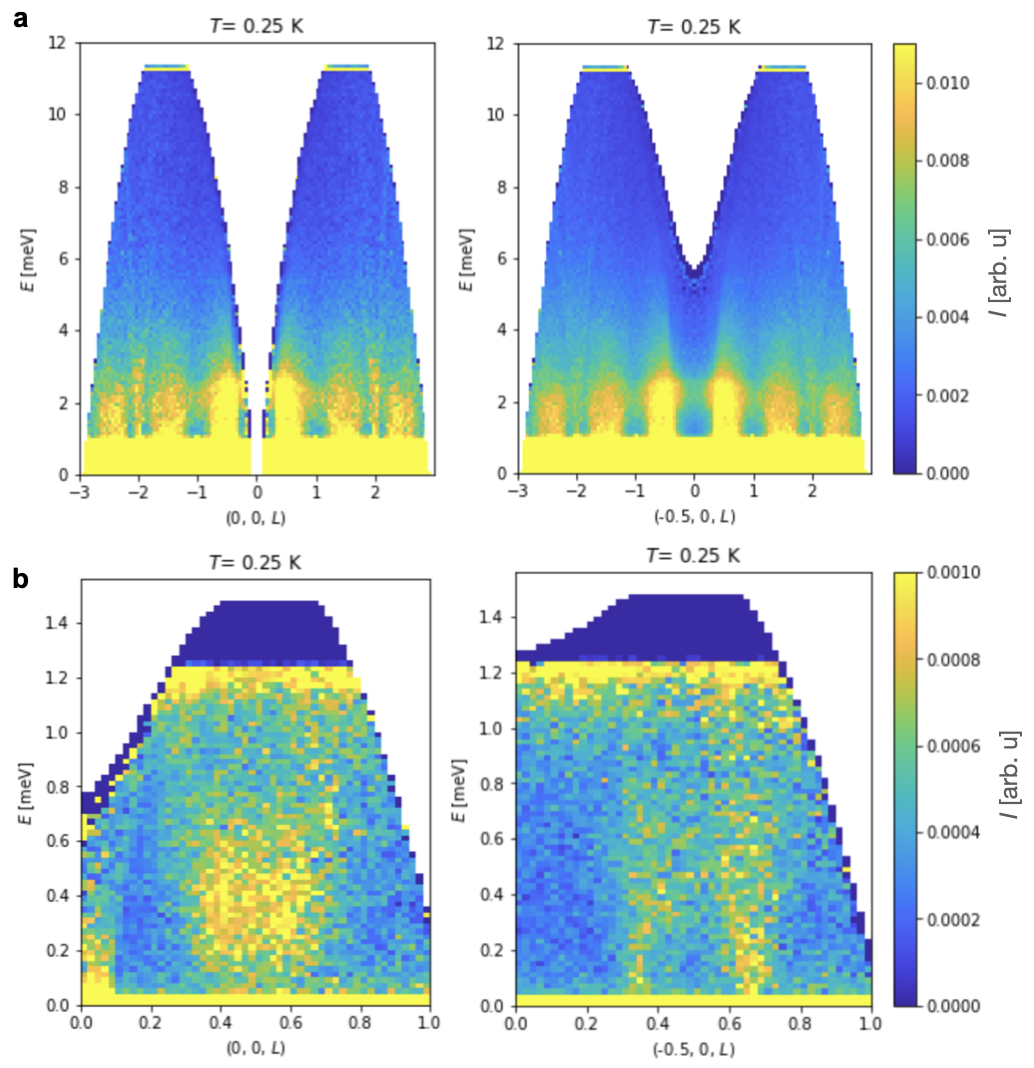}\\
\caption{\label{fig:ins0T}%
\textbf{Magnetic excitations at $\mu_0H$ = 0 T.} Magnetic excitation spectrum of YbAgGe along (0, 0, $L$) and (-1/2, 0, $L$) measured at $T$ = 0.25 K and $E_i$ = 12 and 1.55 meV in \textbf{a} and \textbf{b}, respectively. We used a reciprocal space integration range $\Delta Q$ = [-0.2, 0.2] perpendicular to the cutting direction.}
\end{figure*}

The static and dynamic properties at $\mu_0H$ =  0 T and $T$ = 0.25 - 30 K were studied on CNCS using incident energies $E_i$ = 12, 3.32 and 1.55 meV with elastic energy resolutions $\Delta E$ = 1.2, 0.175 and 0.04 meV FWHM. While we show the $E_i$ = 3.32 meV dataset at $T$ = 0.25 K in Fig. 3 of the main manuscript, the $E_i$ = 12 and 1.55 meV datasets along (0, 0, $L$) and (-1/2, 0, $L$) are shown in Suppl. Fig. \ref{fig:ins0T}a and b, respectively. The $E_i$ = 12 meV dataset provides evidence that the column like excitations at ($H$, 0, 1/3), ($H$, 0, 2/3) and (0, 0, 1/2) are the only observable magnetic excitations below the first CEF transition at an energy transfer $E$ = 12 meV. The additional cone-like signal around (0, 0, 2) is an acoustic phonon. The high-resolution setting at $E_i$ = 1.55 meV reveals no clear excitation gap at low energy transfer. This is further supported by the one-dimensional energy cuts various reciprocal lattice positions shown in Suppl. Fig. \ref{fig:instemp}a, suggesting that the excitations are quasielastic.

\begin{figure*}[tbh]
\centering
\includegraphics[width=\textwidth]{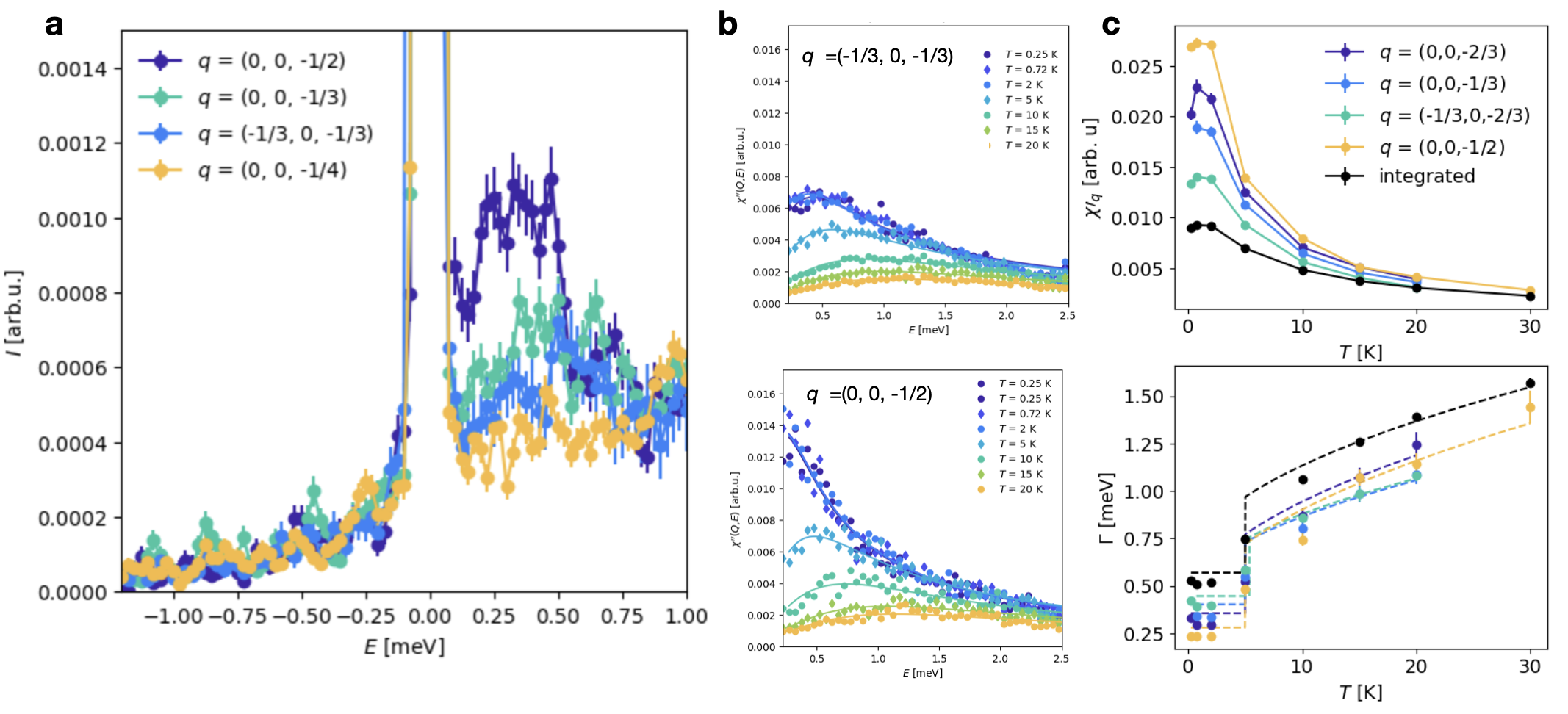}\\
\caption{\label{fig:instemp}%
\textbf{Temperature dependence of the dynamic fluctuations.} \textbf{a} One dimensional line cuts through various reciprocal lattice positions at $T$ = 0.25 K using $E_i$ = 1.55 meV with an integration of $\Delta Q$ = [-0.05, 0.05]. \textbf{b} One dimensional line cuts through $q$ = (0, 0, -1/2) and (-1/3, 0, -1/3) as function of temperature using $E_i$= 3.32 meV with an integration range of $\Delta Q$ = [-0.05, 0.05] around the reciprocal lattice positions. Overplotted (dashed lines) are fitted quasielastic Lorentzian lineshapes  $\chi''$($q$, $E$) = ($E\chi'_q\Gamma_q$)/($E^2$+$\Gamma^2_q$). $\chi''$($q$, $E$) is the imaginary part of the dynamic susceptibility directly revealed by neutron scattering, $\chi'_q$ is real part of the static susceptibility at the wavevector $q$, and $\Gamma_q$ is a characteristic energy scale. The temperature dependence of $\chi'_q$ and $\Gamma_q$ for different $q$s are shown in \textbf{c}. The characteristic energy scale follows the dependence $\Gamma_q(T)$ = $\Gamma_q(0)$  + $\Theta(T-T_s)A\sqrt{T}$ with $A$ = 0.19(2) meVK$^{-1/2}$, and $T_s\approx$ 5 K  occurring in the step function $\Theta$ is close to where  $\chi'_q$ is maximal.}
\end{figure*}

Following reported inelastic neutron scattering experiments at specific ($H$, 0, 1/3) and (1/3, 0, $L$) positions \cite{Fak2005}, we fitted the dynamic susceptibility to a quasielastic Lorentzian $\chi''$($q$, $E$) = ($E\chi'_q\Gamma_q$)/($E^2$+$\Gamma^2_q$). $\chi''$($q$, $E$) is directly probed by neutron scattering, $\chi'_q$ is the real part of the static susceptibility at the wavevector $q$, and $\Gamma_q$ is a characteristic energy scale. The results are shown in Suppl. Fig. \ref{fig:instemp}b,c. We find a wavevector dependent $\Gamma_q$ that is smallest at $q$ = (0, 0, 1/2) (0.23(7) meV), 0.41(6) meV at $q$ = (0, 0, 1/3) and  0.54(7) meV in average. Their finite values attest that no quantum critical fluctuations are obscuring the magnetic excitation spectrum at $\mu_0H$ = 0 T. Upon increasing temperature the quasielastic Lorentzian fit shows that the wavevector dependence of the static susceptibility $\chi'_q$ is suppressed above $T^*$. The characteristic energy scale follows the dependence $\Gamma_q(T)$ = $\Gamma_q(0)$  + $\Theta(T-T_s)A\sqrt{T}$ with $A$ = 0.19(2) meVK$^{-1/2}$, and $T_s\approx$ 5 K  occurring in the step function $\Theta$ is close to where  $\chi'_q$ is maximal.
\section*{\texorpdfstring{S\MakeLowercase{upplementary} N\MakeLowercase{ote 6:}}{Supplementary Note} Magnetic field dependence of the dynamic properties}

\begin{figure*}[tbh]
\centering
\includegraphics[width=\textwidth]{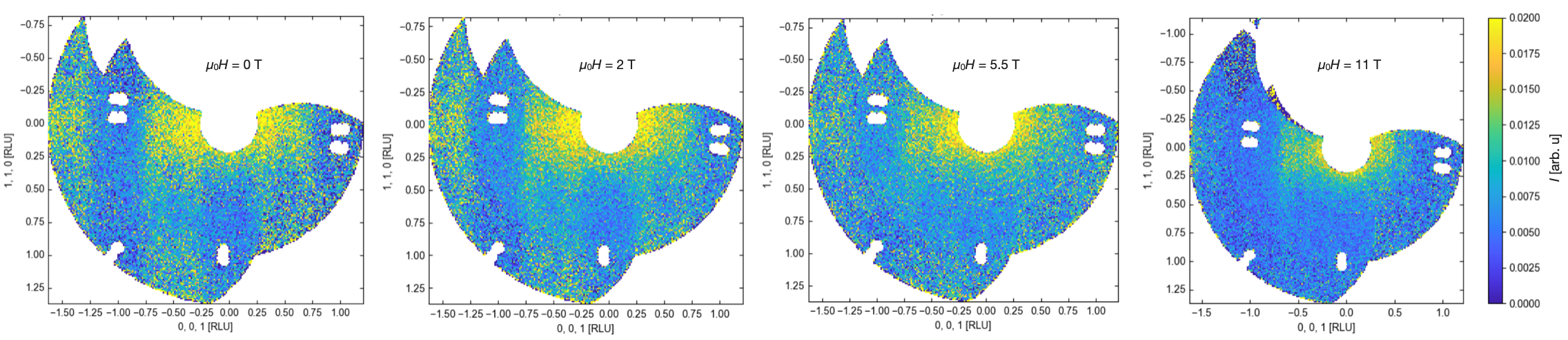}\\
\caption{\label{fig:insmag}%
\textbf{Field dependence of excitations.} Reciprocal space maps of the ($H$, $H$, $L$) plane at an energy transfer $E$ = 0.6 meV at $\mu_0H$ = 0, 2, 5.5, 11 T along $H\parallel$[1-10]. The data were measured with $E_i$ = 5 meV using an energy integration range $\Delta E$ = [0.5, 0.7] meV.}
\end{figure*}

The magnetic field dependent excitations were measured on the cold neutron multiplexing spectrometer CAMEA. In Suppl. Fig. \ref{fig:insmag} we show reciprocal space maps of the ($H$, $H$, $L$) plane at an energy transfer $E$ = 0.6(1) meV at $\mu_0H$ = 0, 2, 5.5, 11 T. The data were measured with an incident energy $E_i$ = 5 meV.  The zero field measurements confirm the column-like excitations with spectral weight along ($H$, 0, 1/3), ($H$, 0, 2/3) and around (0, 0, 1/2) observed on CNCS. Apart of a decrease of the neutron intensity the excitations remain unchanged until $\mu_0H$ = 5.5 T under a magnetic field along $H\parallel$[1-10], despite various changes in the magnetic long-range order \cite{Larsen2021}. At $\mu_0H$ = 11 T we find a broad dispersion along (0, 0, $L$) that is quasielastic at (0, 0, 1/3) and (0, 0, 2/3) and maximal at (0, 0, 1) with an energy transfer $E$ = 4 meV (see Fig. 5 of the main manuscript).

\section*{\texorpdfstring{S\MakeLowercase{upplementary} N\MakeLowercase{ote 7:}}{Supplementary Note} Linear Spin-wave simulations}

We attempted to model the magnetic excitation spectrum of YbAgGe using a minimal local-moment Hamiltonian. The diffraction experiments in Fig. 2 and 3 of the main manuscript reveal that the correlations length along the $c$-axis are largely exceeding the ones in the Kagome plane. The excitation spectrum shows an energy dependence of the fluctuations that takes place mainly along the $L$-direction, for which earlier diffraction results  reveal a robust  $q_z$ $\approx\frac{1}{3}$ component of the magnetic wavevector (see also SI Note1) \cite{Larsen2021}. Thus, we first looked into one-dimensional Heisnberg chain Hamiltonians with interaction parameters that retain the $q_z\approx\frac{1}{3}$ periodicity and with spin-wave boundaries that broadly reflect the excitation energy range of the $\mu_0H$ = 11 T spectrum shown in Fig. 5 of the main manuscript

\begin{equation}
        H=\sum_{ij} J_{cij}{\bf S}_{i}\cdot{\bf S}_{j}+
        A\sum_{i} {\bf S}_{i}^2-
        g\mu_B\mu_0\bf{H} \sum_{i} \bf{S_i}.
        \label{eq:Hamilins}
\end{equation}
The first term describes the exchange couplings along the $c$-axis between spins $\bf S$ on sites $i$ and $j$ with exchange couplings $J_{cij}$. The second term accounts for the anisotropy $A$, and the last term is the Zeeman term with $H||$[1, -1, 0]. We assumed independent chains along the $c$-axis with two nearest-neighbour interactions $J_{c1}$, and $J_{c2}$, whose ratio leads to a magnetic structure with a propagation vector retaining the $q_z\approx\frac{1}{3}$ periodicity. The values were adjusted such that a spin-wave boundary of $E\sim3.5$ meV was obtained at $\mu_0H$ = 11 T, in accordance to the experimental data (see Fig. 5 of the main manuscript). 

\begin{figure*}[tbh]
\centering
\includegraphics[width=\textwidth]{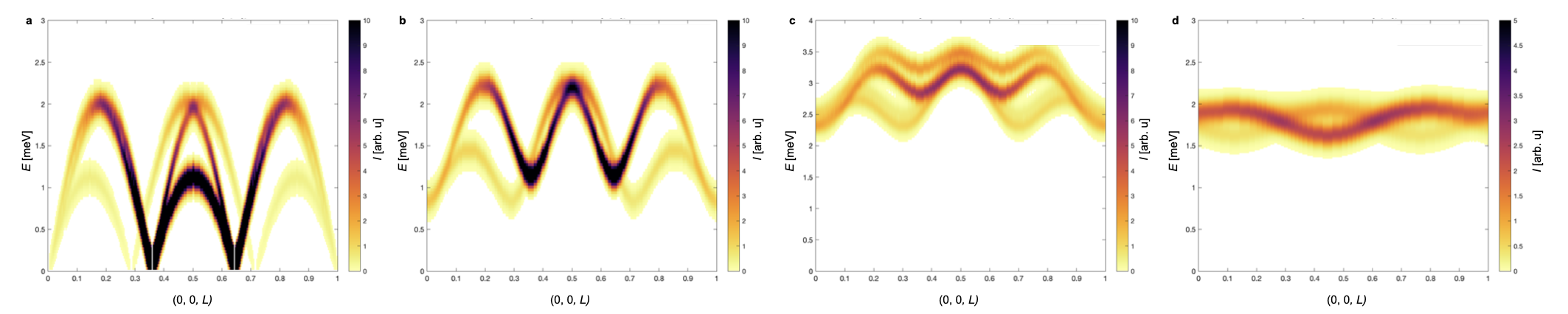}\\
\caption{\label{fig:spinwave}%
\textbf{Linear spin-wave simulation.} Simulation of the magnetic excitation spectrum along (0, 0, $L$) using a one dimensional Heisenberg chain without local anisotropy in \textbf{a}, with a small axial anisotropy in \textbf{b}, and under magnetic field along $H||[1, -1, 0]$ in \textbf{c}. Subpanel \textbf{d} shows a simulation of a $xxz$-chain, in which the crystal-field anisotropy was incorporated via anisotropic exchange interactions.}
\end{figure*}

Supplementary Figure \ref{fig:spinwave}a shows the spin-wave dispersion along (0, 0, $L$) for the Heisenberg chain with antiferromagnetic interactions $J_{c1}$ = 1.5 meV, $J_{c2}$ = 0.6 meV and $A$ = 0 at zero magnetic field. The configurations yield a spiral structure with magnetic Goldstone modes within $E$ = 0 - 2 meV. Both, the finite axial anisotropy $A<0$ mimicking the crystal-field anisotropy and non-zero magnetic field lead to a gaped excitation spectrum, as shown in Suppl. Figs. \ref{fig:spinwave}b, c for $A$ = -0.1 meV and $\mu_0H$ =  11 T, respectively. This contrasts the experimental observations, which report soft excitations up to $\mu_0H$ =  11 T. Thus, we concluded that a one-dimensional Heisenberg chain does not provide a satisfactory description of the observed magnetic excitation spectrum. 
In an alternative approach we incorporated the crystal-field anisotropy via anisotropic exchange interactions using the effective Hamiltonian
\begin{equation}
        H=\sum_{ij} J_{cij}^{xx}S_{i}^x\cdot S_{j}^x + J_{cij}^{yy}S_{i}^y\cdot S_{j}^y+J_{cij}^{zz} S_{i}^z\cdot S_{j}^z
        \label{eq:Hamilins_ani}
\end{equation}
 with $J_{c1}^{xx}$ = $J_{c1}^{yy}$ = 0.2, $J_{c1}^{zz}$ = 1.5, $J_{c2}^{xx}$ = $J_{c2}^{yy}$ = 0 and  $J_{c2}^{zz}$ = 0.6 meV. The model yields a magnetic structure with $q_z\approx\frac{1}{3}$. The resulting spin waves are displayed in Suppl. Fig. \ref{fig:spinwave}d, which is also inconsistent with the experimental observations.

}
\end{document}